\documentclass[aps,prd,twocolumn,nofootinbib,showpacs,superscriptaddress,floatfix]{revtex4-1}

\usepackage{graphicx,enumerate}
\usepackage{amsmath,epsfig}

\newcommand{\be}{\begin{equation}}
\newcommand{\ee}{\end{equation}}
\newcommand{\ba}{\begin{eqnarray}}
\newcommand{\ea}{\end{eqnarray}}
\newcommand{\nn}{\nonumber}

\begin{document}

\title{X-ray probes of black hole accretion disks for testing the no-hair theorem}

\author{Tim Johannsen}
\affiliation{Department of Physics and Astronomy, University of Waterloo, Waterloo, Ontario, N2L 3G1, Canada}
\affiliation{Canadian Institute for Theoretical Astrophysics, University of Toronto, Toronto, Ontario M5S 3H8, Canada}
\affiliation{Perimeter Institute for Theoretical Physics, Waterloo, Ontario, N2L 2Y5, Canada}

\begin{abstract}

The spins of a number of supermassive and stellar-mass black holes have been measured based on detections of thermal continuum emission and relativistically broadened iron lines in their x-ray spectra. Likewise, quasiperiodic variability has been observed in several sources. Such measurements commonly make the assumption that black holes are described by the Kerr metric, which according to the no-hair theorem characterizes black holes uniquely in terms of their masses and spins. This fundamental property of black holes can be tested observationally by measuring potential deviations from the Kerr metric introduced by a parametrically deformed Kerr-like spacetime. Thermal spectra, iron lines, and variability have already been studied extensively in several such metrics, which usually depend on only one particular type of deviation or contain unphysical regions outside of the compact object. In this paper, I study these x-ray probes in the background of a new Kerr-like metric which depends on four independent deviation functions and is free of pathological regions outside of the event horizon. I show that the observed signals depend significantly on primarily two types of deviations and that the strong correlation between the spin and the deviation parameters found previously in other Kerr-like metrics is partially broken for rapidly spinning black holes. This suggests that high-spin sources are the best candidates for tests of the no-hair theorem with x-rays and I obtain first constraints on such deviations from the stellar-mass black hole Cygnus~X--1.

\end{abstract}


\maketitle

%%%%%%%%%%%%%%%%%%%%%%%%%%%%%%%%%%%%%%%%%%
\section{INTRODUCTION}

Black holes are encountered ubiquitously in the centers of galaxies and in x-ray binaries \cite{kor95,MR06}. In general relativity, stationary black holes in vacuum are fully characterized by the no-hair theorem which states that they only depend on two parameters, their masses $M$ and spins $J$, and are described by the Kerr metric (see, e.g., Ref.~\cite{heu96}). In practice, such black holes will not be fully stationary or exist in perfect vacuum due to the presence of, e.g., surrounding matter or other stars, but it is often reasonable to assume that these effects are sufficiently small to be negligible. While observational evidence exists for the presence of event horizons (e.g., \cite{nar97}), a final proof of the Kerr nature of these compact objects is still lacking.

General relativity has been well confirmed in the regime of weak spacetime curvatures \cite{will06} but still remains practically untested in the strong-field regime that is found around compact objects (see Ref.~\cite{psalLRR}). Indeed, black holes need not necessarily be described by the Kerr metric at all (e.g., \cite{pani09}). Since it is unclear at present whether general relativity is modified in the strong-field regime and, if so, in what manner, an efficient approach is to test the no-hair theorem in a model-independent manner that is based on a parametrically deformed Kerr-like metric which encompasses many different theories of gravity at once. Such metrics generally do not derive from the action of any particular gravity theory, and the underlying theory is usually unknown. However, it is assumed that particles follow the geodesics of the spacetime.

Several such metrics have been constructed thus far (e.g., \cite{MN92,CH04,GB06,VH10,JPmetric,vig11}), which depend on one or more free parameters that measure potential deviations from the Kerr metric and which include the Kerr metric as the special case when all deviations vanish. Observations can then be used to measure these deviations, should they exist, and, thereby, infer properties of the underlying but generally unknown theory of gravity. If no deviations are detected, the compact object is verified to be a Kerr black hole. If, on the other hand, nonzero deviations are measured, there are two possible interpretations. If general relativity still holds, the object is not a black hole but, instead, another stable stellar configuration or, perhaps, an exotic object \cite{Hughes06}. Otherwise, the no-hair theorem would be falsified.

Tests of the no-hair theorem have been suggested based on either the gravitational-wave signals of extreme mass-ratio inspirals (see Ref.~\cite{gair13} for a review) or electromagnetic observations of pulsar black hole binaries \cite{pulsars}, stars on orbits around Sgr~A* \cite{Sgr}, or of the radiation from the accretion flows of black holes (e.g., \cite{PaperI}). The emission from black hole accretion disks provides a number of different observables that could be used for such tests, including the direct imaging of the shadows of supermassive black holes with very-long baseline interferometric arrays (e.g., \cite{joh10b,bamyosh10,joh13rings,bro13}), relativistically broadened iron lines (e.g., \cite{jp13,bambiiron,joh13edges}), thermal disk spectra (e.g., \cite{bambar11,bambidiskcode,bambidiskjet,krawc12}), quasiperiodic variability \cite{joh11a,jp13,bambiqpo}, and x-ray polarization \cite{krawc12}.

Previous model-independent studies in the electromagnetic spectrum were largely based on either a quasi-Kerr metric \cite{GB06} or a metric proposed by Ref.~\cite{JPmetric}. These metrics depend on only one type of potential deviations from the Kerr metric, although there are at least four independent ways to parametrize them (cf. \cite{vig11}). Both of these metrics generally harbor naked singularities which makes them less favorable candidates as an alternative to black holes. In addition, the quasi-Kerr metric, just as many other Kerr-like metrics, contains pathological regions near the central object where other singularities occur or causality is violated. Therefore, an artificial cutoff needs to be imposed in both metrics which shields the singularities and any unphysical region from outside observers. In this form, then, these metrics are valid frameworks for tests of the no-hair theorem (see Ref.~\cite{joh13pathols} for a full discussion). Note, however, that the metric of Ref.~\cite{JPmetric} uses the Newman-Janis algorithm~\cite{NewmanJanis} to generate rotating solutions from static seeds and it is not guaranteed that this procedure can be applied consistently to general metrics which are not solutions of the Einstein field equations.

Under the assumption that black holes are described by the Kerr metric, thermal disk spectra and relativistically broadened iron lines have already been widely used to measure the spins of stellar-mass black holes and active galactic nuclei (AGNs); see Refs.~\cite{mccl13,rey13} for recent reviews. The latter method assumes irradiation off the accretion disks of black holes in the form of fluorescent iron lines which appear significantly broadened due to the relativistic effects of light bending, Doppler boosting and beaming, and the gravitational redshift \cite{ironmethod,dex09}. The former method is based on the Novikov-Thorne model \cite{nov73}, a relativistic thin-disk accretion flow model, which describes the emission of thermal radiation. Several authors have computed the observed spectra with the inclusion of the relativistic effects \cite{spectra,cunn76,li05}. The current state-of-the-art fitting routine combines such spectra with a suitable model of the disk atmosphere to account for spectral hardening \cite{mccl06}. Both of these methods aim to measure the location of the innermost stable circular orbit (ISCO) which depends directly on the spin of the black hole.

Quasiperiodic oscillations (QPOs) have been observed in both stellar-mass black holes and AGNs \cite{MR06,gier08} and provide a potential third means to measure their spins. QPOs are transient and occur mostly in nonthermal accretion disk states and during state transitions. They can be divided into two general classes: high-frequency QPOs (roughly 40--450~Hz) and low-frequency QPOs (roughly 0.1--30~Hz) and their observed frequencies sometimes fall into ratios of small integers; see Ref.~\cite{MR06} for a detailed discussion. A number of different QPO models have been designed to date (e.g., \cite{per97,otherQPOmodels,sil01,kluz01,abra03}), but at present the exact origin of QPOs is still unknown.

In this paper, I analyze thermal disk spectra, relativistically broadened iron lines, and quasiperiodic variability in the context of a new Kerr-like metric \cite{joh13metric}. This metric describes a black hole and is free of singularities or other adverse regions outside of the event horizon. Likewise, this metric depends on four independent deviation functions which parametrize possible departures from the Kerr metric. I show that these observables depend primarily on two deviation types which can lead to significantly modified signals that should be resolvable with either current or near-future x-ray missions.

While the spin can be inferred uniquely from the location of the ISCO for Kerr black holes, in Kerr-like metrics the ISCO generally depends on the spin and the deviation parameters (e.g., \cite{PaperI}). References~\cite{PJ12,jp13,joh13edges} showed that iron line profiles are practically indistinguishable for values of the spin and the deviation parameter that correspond to the same ISCO in either the quasi-Kerr metric \cite{GB06} or the metric of Ref.~\cite{JPmetric}. Reference~\cite{bambidiskcode} found a similar result for thermal disk spectra obtained in the latter metric. Thus, there is a strong correlation between the spin and the respective deviation parameters which hampers the ability of these observables (as well as of quasiperiodic variability) to clearly measure such deviations if they exist.

I demonstrate that both the thermal spectra and iron line profiles are also very similar in the new Kerr-like metric for values of the spin and the deviation parameters with the same ISCO if the ISCO is located at comparatively large radii corresponding to Kerr black holes with small to intermediate spins. However, for ISCO radii that are sufficiently close to the event horizon, i.e., for rapidly spinning black holes, this strong correlation among the spin and the deviation parameters is partially broken and the thermal spectra and line profiles become distinguishable. I also show that a similar property holds for the QPOs predicted by the diskoseismology (see Ref.~\cite{wag08} for a review) and resonance \cite{kluz01,abra03} models.

This suggests that black holes with high spins are the best candidates for tests of the no-hair theorem with x-ray observables. Based on the recent near-maximal spin measurement of the Galactic black hole Cygnus~X--1 from its thermal spectrum by Gou et al.~\cite{Gou14}, I obtain first constraints on two deviation parameters of the new metric.

In Sec.~\ref{sec:metric}, I summarize some of the properties of the Kerr and Kerr-like metrics. In Sec.~\ref{sec:code}, I describe the ray tracing algorithm that I use to simulate the observed spectra and line profiles. In Secs.~\ref{sec:diskspectra}, \ref{sec:ironlines}, and \ref{sec:qpo}, I describe the effects of the deviation parameters on the thermal spectra, iron lines, and QPOs, respectively. I discuss my conclusions in Sec.~\ref{sec:conclusions}. From here on, I use geometric units and set $G=c=1$, where $G$ and $c$ are the gravitational constant and the speed of light, respectively, unless otherwise stated. In these units, length (and time) scales are expressed in terms of the mass of the black hole, where $M=GM/c^2\equiv r_g$, the gravitational radius.

%%%%%%%%%%%%%%%%%%%%%%%%%%%%%%%%%%%%%%
\section{KERR AND KERR-LIKE BLACK HOLE METRICS}
\label{sec:metric}

In this section, I describe some of the properties of black holes that are described by the two metrics I consider in this paper. Expressed in Boyer-Lindquist coordinates, these are the Kerr metric $g_{\mu\nu}^{\rm K}$,
\ba
g_{tt}^{\rm K} &=&-\left(1-\frac{2Mr}{\Sigma}\right), \nonumber \\
g_{t\phi}^{\rm K} &=& -\frac{2Mar\sin^2\theta}{\Sigma}, \nonumber \\
g_{rr}^{\rm K} &=& \frac{\Sigma}{\Delta}, \nonumber \\
g_{\theta \theta}^{\rm K} &=& \Sigma, \nonumber \\
g_{\phi \phi}^{\rm K} &=& \left(r^2+a^2+\frac{2Ma^2r\sin^2\theta}{\Sigma}\right)\sin^2\theta,
\label{eq:kerr}
\ea
where
\ba
\Delta &\equiv& r^2-2Mr+a^2, \nn \\
\Sigma &\equiv& r^2+a^2\cos^2 \theta,
\label{eq:deltasigma}
\ea
and the Kerr-like metric of Ref.~\cite{joh13metric},
\ba
g_{tt} &=& -\frac{\tilde{\Sigma}[\Delta-a^2A_2(r)^2\sin^2\theta]}{[(r^2+a^2)A_1(r)-a^2A_2(r)\sin^2\theta]^2}, \nn \\
g_{t\phi} &=& -\frac{a[(r^2+a^2)A_1(r)A_2(r)-\Delta]\tilde{\Sigma}\sin^2\theta}{[(r^2+a^2)A_1(r)-a^2A_2(r)\sin^2\theta]^2}, \nn \\
g_{rr} &=& \frac{\tilde{\Sigma}}{\Delta A_5(r)}, \nn \\
g_{\theta \theta} &=& \tilde{\Sigma}, \nn \\
g_{\phi \phi} &=& \frac{\tilde{\Sigma} \sin^2 \theta \left[(r^2 + a^2)^2 A_1(r)^2 - a^2 \Delta \sin^2 \theta \right]}{[(r^2+a^2)A_1(r)-a^2A_2(r)\sin^2\theta]^2},
\label{eq:metric}
\ea
where
\ba
A_1(r) &=& 1 + \sum_{n=3}^\infty \alpha_{1n} \left( \frac{M}{r} \right)^n, 
\label{eq:A1}\\
A_2(r) &=& 1 + \sum_{n=2}^\infty \alpha_{2n} \left( \frac{M}{r} \right)^n, 
\label{eq:A2}\\
A_5(r) &=& 1 + \sum_{n=2}^\infty \alpha_{5n} \left( \frac{M}{r} \right)^n, 
\label{eq:A5}\\
\tilde{\Sigma} &=& \Sigma + f(r), 
\label{eq:Sigmatilde}\\
f(r) &=& \sum_{n=3}^\infty\epsilon_n \frac{M^n}{r^{n-2}}.
\label{eq:f}
\ea
In these expressions, $a\equiv J/M$ is the spin parameter.

The latter metric contains the four free functions $f(r)$, $A_1(r)$, $A_2(r)$, and $A_5(r)$ that depend on four sets of parameters which measure potential deviations from the Kerr metric. In the case when all deviation parameters vanish, i.e., when $f(r)=0$, $A_1(r)=A_2(r)=A_5(r)=1$, this metric reduces to the Kerr metric in Eq.~(\ref{eq:kerr}). 

Just as the Kerr metric, the metric in Eq.~(\ref{eq:metric}) is stationary, axisymmetric, and asymptotically flat. Unlike the Kerr metric, which is a vacuum solution of the Einstein equations, the metric in Eq.~(\ref{eq:metric}) is not a vacuum solution in general relativity, because it describes black holes in modified theories of gravity that differ from Kerr black holes. Thanks to its structure, the Kerr-like metric in Eq.~(\ref{eq:metric}) is phenomenological; i.e., it encompasses large classes of modified gravity theories and can be mapped to the known black hole solutions of particular theories for certain choices of the deviation functions including Chern-Simons gravity and Einstein-Dilaton-Gauss-Bonnet gravity (see the appendix of Ref.~\cite{joh13metric} for the respective mappings). Using this metric as a framework, one can then test observationally whether astrophysical black holes are indeed described by the Kerr metric.

In general relativity, stationary, axisymmetric, and asymptotically flat metrics that admit the existence of integrable two-dimensional hypersurfaces generally depend on only four functions. As a consequence of Frobenius's theorem, such hypersurfaces are automatically guaranteed to exist if the spacetime is also vacuum. Such metrics can be written in the form of the Papapetrou line element where the metric is expressed with respect to the Weyl-Papapetrou coordinates and has only three metric functions. Out of these functions only two are independent, while the third one can be derived from the other two.

This happens for three reasons. First, the symmetries imposed, the assumption of asymptotic flatness and the vanishing of the Ricci tensor allow for the spacetime to have integrable two-dimensional hypersurfaces that are orthogonal to the two Killing fields and on which one can define coordinates that can be
carried along integral curves of these Killing fields to the rest of the spacetime. Thus the metric can be written in a $2\times2$ block form. If one chooses as one of the coordinates the determinant of the $(t,\phi)$ part of the metric, then the metric can be written in a form that has only four independent functions. Second, the field equations imply, by the vanishing of the Ricci tensor (i.e., the vacuum assumption), that the coordinate $\rho$ which is defined by the determinant of the $(t,\phi)$ part of the metric is a harmonic function and thus one can define the second coordinate on the two-dimensional hypersurfaces as the harmonic conjugate of $\rho$ and absorb one of the functions in the process, reducing the independent functions to three. Finally the vacuum field equations imply that the third of the functions is related to the other two and can be determined up to the addition of a constant \cite{Wald84}.

In alternative theories of gravity, however, the vacuum assumption does not necessarily imply the existence of two-dimensional integrable hypersurfaces. Therefore, one would expect that metrics which describe black holes in alternative theories of gravity depend on at least four independent functions which motivates the introduction of four deviation functions in Eq.~(\ref{eq:metric}). The quasi-Kerr metric \cite{GB06} and the metric of Ref.~\cite{JPmetric} depend on only one such function making them less general in this sense. Nonetheless, these metrics can likewise be used for observational tests of the nature of black holes. Recently, Cardoso et al. \cite{CPR14} generalized the metric of Ref.~\cite{JPmetric} to include two deviation types.

The deviation functions in Eqs.~(\ref{eq:A1})--(\ref{eq:f}) are written as power series in $M/r$. The lowest-order coefficients of these series vanish so that the deviations from the Kerr metric are consistent with all current weak-field tests of general relativity (see Ref.~\cite{will06}). In this paper, I will focus on black holes that are described by the lowest-order metric, which only depend on the mass $M$, spin $a$, and the deviation parameters $\epsilon_3$, $\alpha_{13}$, $\alpha_{22}$, and $\alpha_{52}$.

Certain lower limits on the deviation parameters $\epsilon_3$, $\alpha_{13}$, $\alpha_{22}$, and $\alpha_{52}$ exist which are determined by the properties of the event horizon. In order for an event horizon to be present these parameters must obey the relations
\ba
\epsilon_3 &>& B_3,~~~~~\alpha_{13} > B_3, \nn \\
\alpha_{22} &>& B_2,~~~~~\alpha_{52} > B_2,
\ea
where
\ba
B_2 &\equiv& - \frac{\left(M+\sqrt{M^2-a^2}\right)^2}{M^2}, \nn \\
B_3 &\equiv& - \frac{\left(M+\sqrt{M^2-a^2}\right)^3}{M^3}.
\label{eq:lowerbounds}
\ea
For values of the deviation parameters smaller than these limits the metric harbors a naked singularity instead of a black hole \cite{joh13metric}. The event horizon itself is independent of these parameters and coincides with the event horizon of a Kerr black hole,
\be
r_+ \equiv M+\sqrt{M^2-a^2}.
\ee
As discussed in Refs.~\cite{joh13metric,joh13rings} the deviation parameters affect the locations of the circular photon orbit and ISCO, as well as the energy, axial angular momentum, and dynamical frequencies of particles on circular equatorial orbits. For nonzero values of the deviation parameters the location of the ISCO is shifted and the orbital velocity of the disk particles either increases or decreases \cite{joh13metric} (cf. Refs.~\cite{PaperI,JPmetric}). In addition, photons which are emitted by an accretion disk surrounding the black hole and which are detected by a distant observer also experience a modified gravitational redshift, relativistic boosting and beaming, as well as light bending (cf. Refs.~\cite{PaperI,jp13}). These properties can have a significant effect on the observed spectra and QPOs (see \cite{jp13,bambidiskcode}).

Both the Kerr and the Kerr-like metrics admit three constants of motion thanks to their symmetries \cite{cart68,joh13metric}. This means that for a particle with rest mass $\mu$ and 4-momentum $p^\alpha$ on a geodesic orbit its energy $E$, angular momentum $L_z$ about the rotation axis of the black hole, and the so-called Carter constant are conserved. The Carter constant has the same form in both metrics, given by the expression
\be
Q \equiv p_\theta^2 - (L_z-aE)^2 + \mu^2a^2\cos^2\theta + \frac{1}{\sin^2\theta}(L_z-aE\sin^2\theta)^2.
\label{eq:CarterQ}
\ee

While in general relativity the no-hair theorem establishes that the Kerr metric is the unique black hole metric (assuming the black hole is electrically neutral), in alternative theories of gravity a Kerr-like metric generally possesses only two constants of motion which correspond to its stationarity and axisymmetry. Therefore, the choice of the metric in Eq.~(\ref{eq:metric}) necessarily restricts potential deviations from the Kerr metric to belong to the class of Kerr-like metrics with three constants of motion and it would be desirable to employ an even more general Kerr-like metric which contains all Kerr-like metrics with at least two constants of motion. At present, however, no such metric is known.

On the other hand, the existence of twodimentional integrable hypersurfaces of the metric in Eq.~(\ref{eq:metric}) has two distinct advantages. First, the known black hole solutions of alternative theories of gravity (all of which can be mapped to this metric~\cite{joh13metric}) possess three constants of motion at least in the case of slowly rotating black holes \cite{nonGRBHs}. Second, the existence of three constants of motion is of great practical use and simplifies many analytic and numerical computations. I will make use of this property in the following.

%%%%%%%%%%%%%%%%%%%%%%%%%%%%%%%%%%%%%%%%%%%%%%%%%
\section{RAY TRACING ALGORITHM}
\label{sec:code}

The existence of three constants of motion can simplify the ray tracing computations required for the study of the emission from the accretion disks of black holes. Thanks to these constants, all four photon equations of motion separate and can be written in first-order form. For the Kerr-like metric in Eq.~(\ref{eq:metric}), these are given by the equations \cite{joh13metric}
\ba
\tilde{\Sigma} \frac{dt}{d\lambda} &=& a(\xi-a\sin^2\theta) + \frac{(r^2+a^2)A_1(r)}{\Delta}P(r),
\label{eq:tEOM} \\
\tilde{\Sigma} \frac{dr}{d\lambda} &=& \pm \sqrt{ A_5(r) R(r) },
\label{eq:rEOM} \\
\tilde{\Sigma} \frac{d\theta}{d\lambda} &=& \pm \sqrt{ \Theta(\theta) },
\label{eq:thetaEOM} \\
\tilde{\Sigma} \frac{d\phi}{d\lambda} &=& \frac{\xi}{\sin^2\theta}-a + \frac{aA_2(r)}{\Delta}P(r),
\label{eq:phiEOM}
\ea
where
\ba
P(r) &\equiv& (r^2+a^2)A_1(r)-aA_2(r)\xi,
\label{eq:P} \\
R(r) &\equiv& P(r)^2 - \Delta \left[ (\xi-a)^2 + \eta \right],
\label{eq:R} \\
\Theta(\theta) &\equiv& \eta + a^2\cos^2\theta - \xi^2\cot^2\theta.
\label{eq:Theta}
\ea
In these expressions, $\lambda$ is an affine parameter, and I used the standard abbreviations
\ba
\xi &\equiv& \frac{L_z}{E},
\label{eq:xi} \\
\eta &\equiv& \frac{Q}{E^2}.
\label{eq:eta}
\ea

Equivalently, these equations can be written in an integral form, which contains only elliptic integrals in the case of the Kerr metric. These can then be calculated elegantly and fast \cite{rau94,dex09}. For the Kerr-like metric in Eq.~(\ref{eq:metric}), however, the equations of motion are significantly more complicated and cannot be expressed in terms of simple elliptic integrals. Therefore, the integral form of these equations does not lead to an increase in computation speed and I integrate the equations of motion instead.

However, a direct integration of Eqs.~(\ref{eq:tEOM})--(\ref{eq:phiEOM}) suffers from an unacceptable loss of precision around the radial and polar turning points where the radial or polar momenta vanish, respectively. Therefore, I follow the scheme used by Ref.~\cite{PJ12} who solved the first-order equations for the coordinates $t$ and $\phi$ and the second-order geodesic equations for $r$ and $\theta$.

As in Ref.~\cite{PJ12}, I consider an accretion disk that lies in the equatorial plane of the black hole with a radial extent from the ISCO to some outer radius $r_{\rm out}$ and that is viewed by an observer at an inclination angle $i$ at a large distance $r_0$ from the black hole. This observer defines an image plane, where photons emitted from the disk are observed. This image plane can be set up in the local rest frame of the observer and has the Cartesian coordinate axes \cite{bar73,joh13rings}
\ba
x' = -r_0 \frac{\xi}{\sqrt{g_{\phi\phi}}\zeta \left( 1 + \frac{g_{t\phi}}{g_{\phi\phi}}\xi \right) },
\label{eq:xprime} \\
y' = r_0 \frac{\pm \sqrt{\Theta(i)} }{\sqrt{g_{\theta\theta}}\zeta \left( 1 + \frac{g_{t\phi}}{g_{\phi\phi}}\xi \right) }.
\label{eq:yprime}
\ea
In the limit $r_0\rightarrow\infty$, these expressions reduce to the simple form (cf. \cite{bar73})
\ba
x' &=& -\frac{\xi}{\sin i},
\label{eq:xprime2} \\
y' &=& \pm \sqrt{\Theta(i)}.
\label{eq:yprime2}
\ea
Note that the $y'$ axis is chosen in a manner such that it lies in a plane with the rotation axis of the black hole and that the origin of the image plane is centered on the black hole.

Since the quantities $\xi$ and $\eta$ are constant along the photon trajectories, I can calculate them at any given point. For each photon corresponding to a point $(x',y')$ in the image plane, I invert Eqs.~(\ref{eq:xprime}) and (\ref{eq:yprime}) to obtain the corresponding constants of motion $\xi$ and $\eta$ and calculate the initial location and 4-momentum of the photon expressed in the coordinate frame centered on the black hole (cf. \cite{PJ12}). I then integrate the equations of motion (\ref{eq:tEOM})--(\ref{eq:phiEOM}) backwards in time until the photon reaches the equatorial plane, where I specify the appropriate boundary conditions for the emissivity depending on the application as I describe in Secs.~\ref{sec:diskspectra} and \ref{sec:ironlines}.

The integrability of the equations of motion can also be used to calculate analytically the observed size of the accretion disk in the image plane. The equations of motion, then, only have to be integrated for image points that lie inside of this region, because only photons corresponding to these points reach the equatorial plane in the radial interval between the ISCO and the outer disk radius. Since the functions $R(r)$ and $\Theta(\theta)$ depend of the constants of motion $\xi$ and $\eta$, this set of image points can be obtained from the requirement that these functions be non-negative. These points form a disk in the image plane, which for large disks extends to a radius that is only slightly larger than $r_{\rm out}$ (cf. \cite{Speith93}). Thanks to the gravitational attraction of the black hole, it is likewise sufficient to integrate the equations of motion only for photons with image points at a radius that is larger than the ISCO.

%%%%%%%%%%%%%%%%%%%%%%%%%%%%%%%%%%%%%%%
\section{THERMAL DISK SPECTRA}
\label{sec:diskspectra}

In this section, I summarize the relevant properties of the thin-disk accretion flow model that I use and calculate the observed thermal disk flux density. I proceed to analyze the effects of the deviation parameters on both the emitted and the observed flux densities.

%%%%%%%%%%%%%%%%%%%%%%%%%%%%%%%%%%%%%%%
\subsection{Thin accretion disk model}
\label{sec:NTdisk}

\begin{figure}[ht]
\begin{center}
\psfig{figure=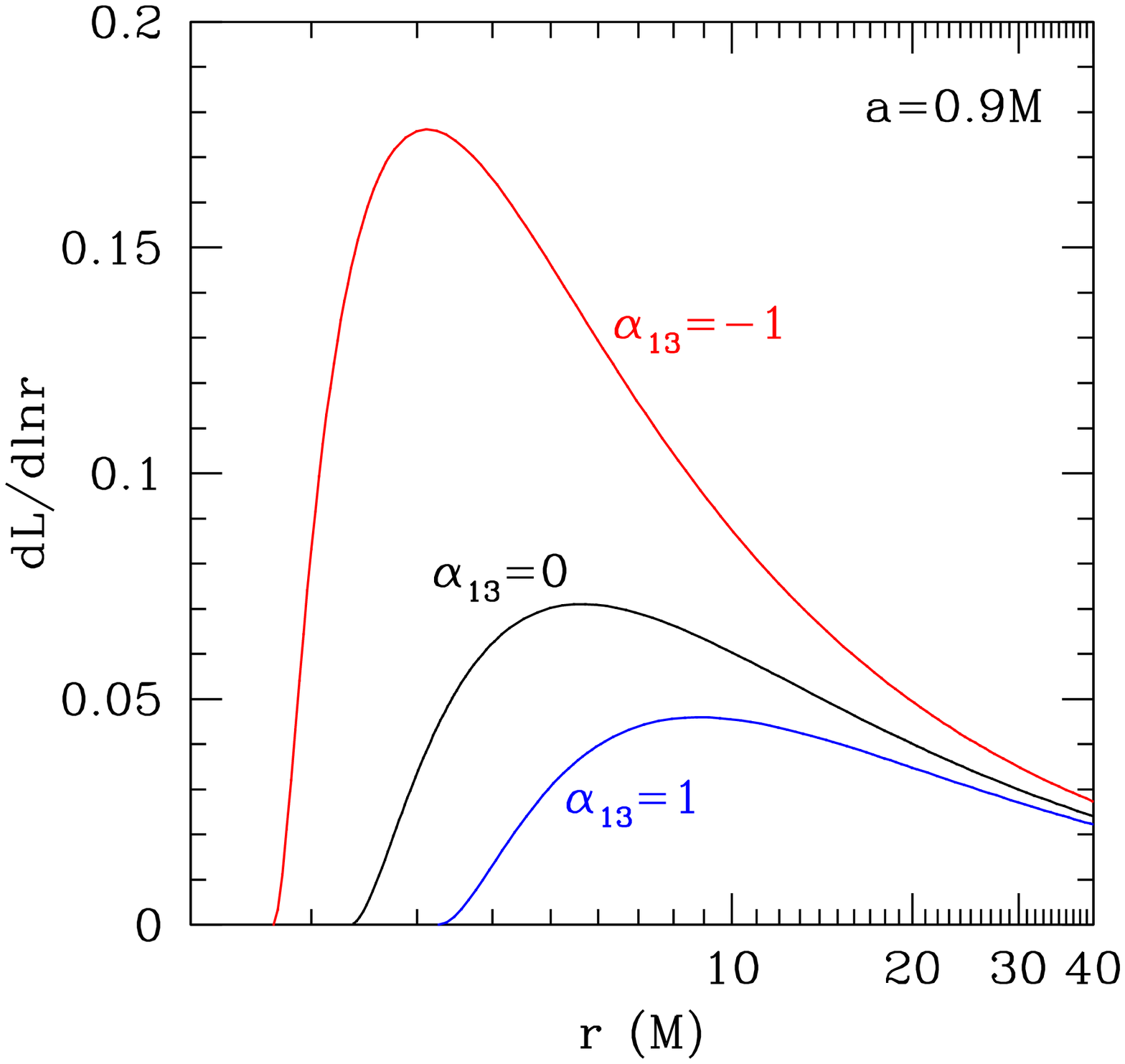,width=0.4\textwidth}
\psfig{figure=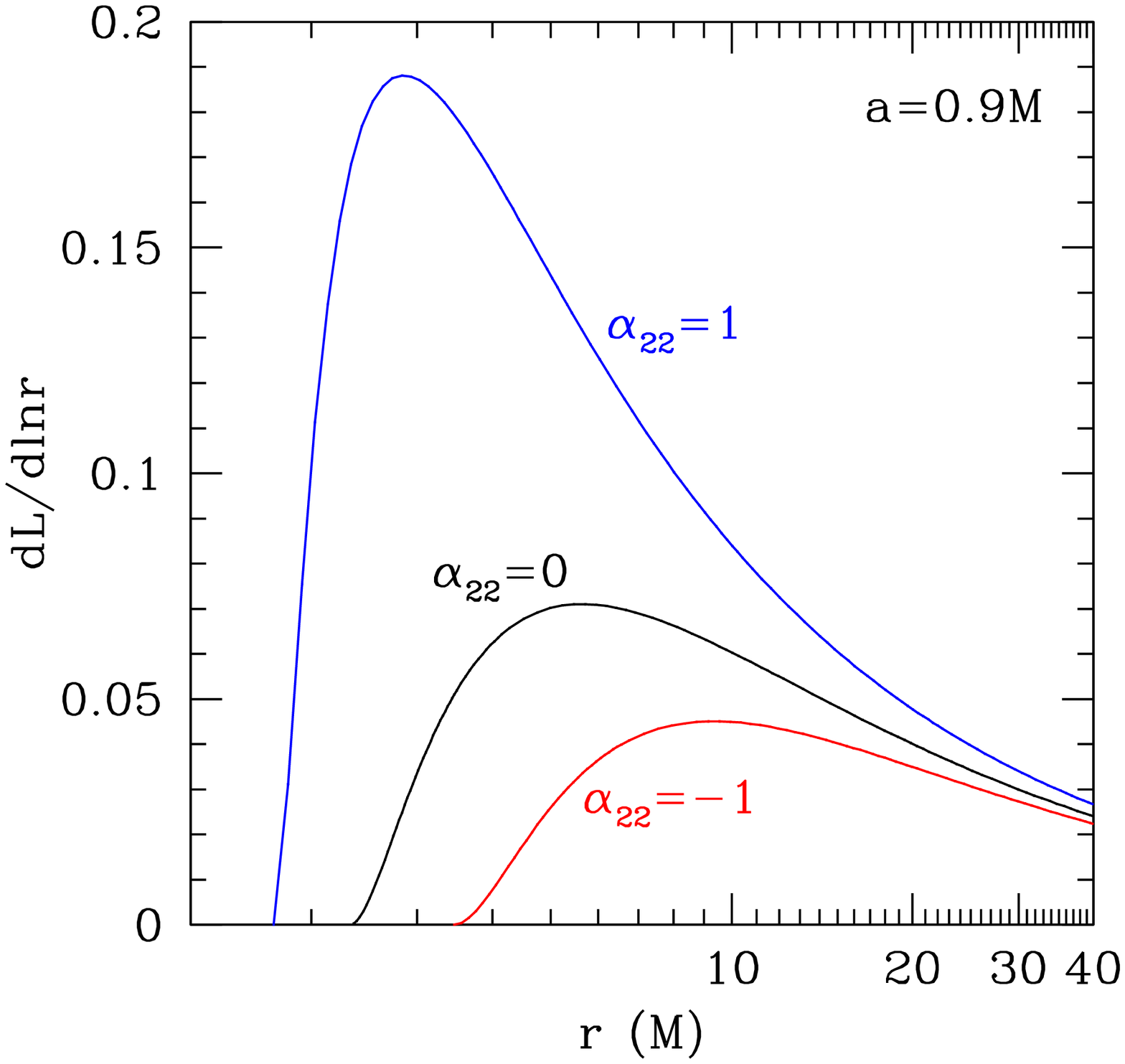,width=0.4\textwidth}
\end{center}
\caption{Radial profiles of the luminosity of an accretion disk (in units of $c^6/4\pi G^2$) around a black hole with spin $a=0.9M$ for several values of the deviation parameters $\alpha_{13}$ (top) and $\alpha_{22}$ (bottom). In each panel, only one deviation parameter is varied, while the other one is set to zero. The luminosity profile is strongly peaked at a radius near the ISCO and decreases rapidly at larger and smaller disk radii. For decreasing values of the parameters $\alpha_{13}$ as well as for increasing values of the parameter $\alpha_{22}$, the peak of the luminosity profile increases and is located closer to the ISCO radius.}
\label{fig:emitflux}
\end{figure}

\begin{figure*}[ht]
\begin{center}
\psfig{figure=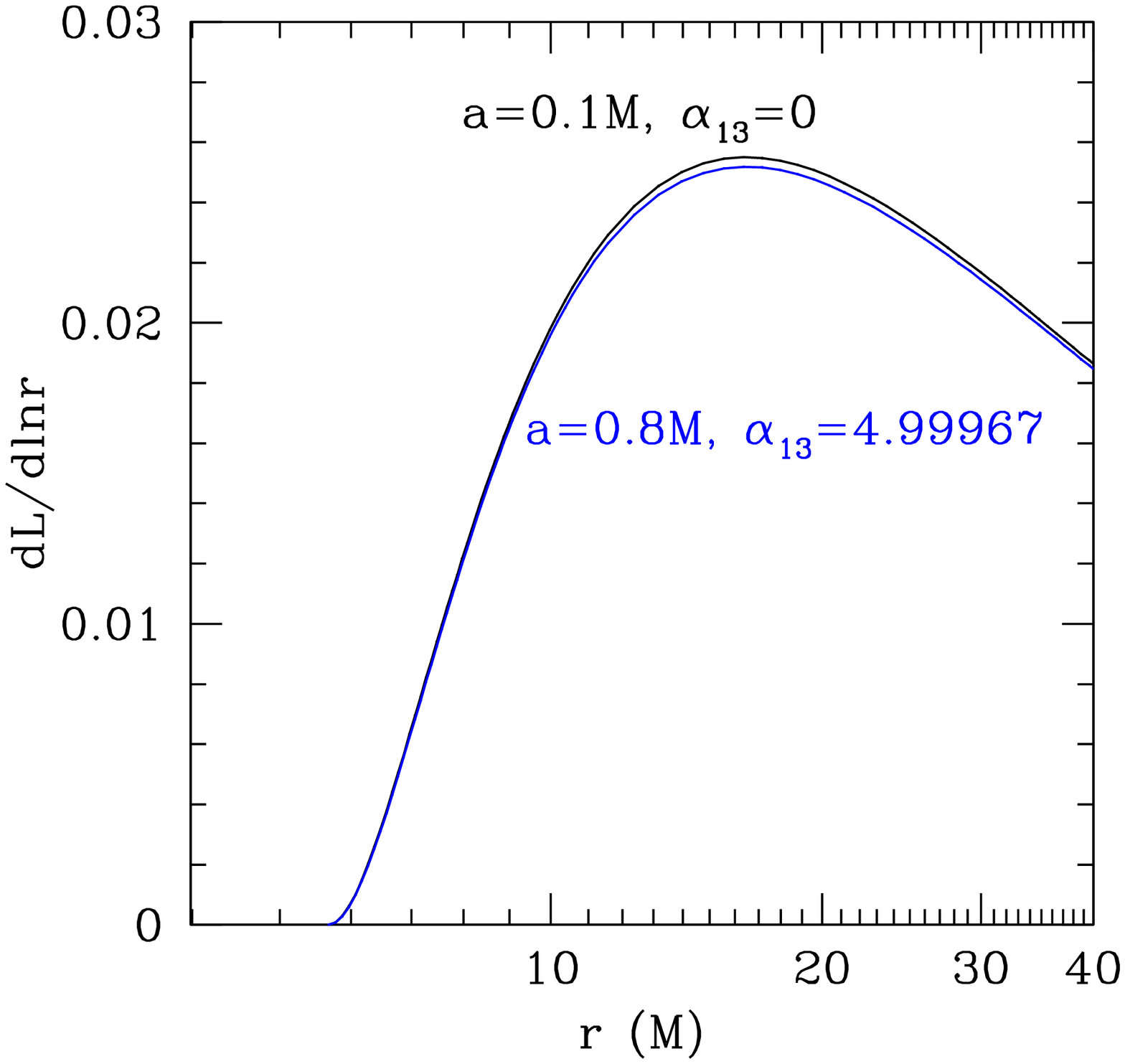,height=1.72in}
\psfig{figure=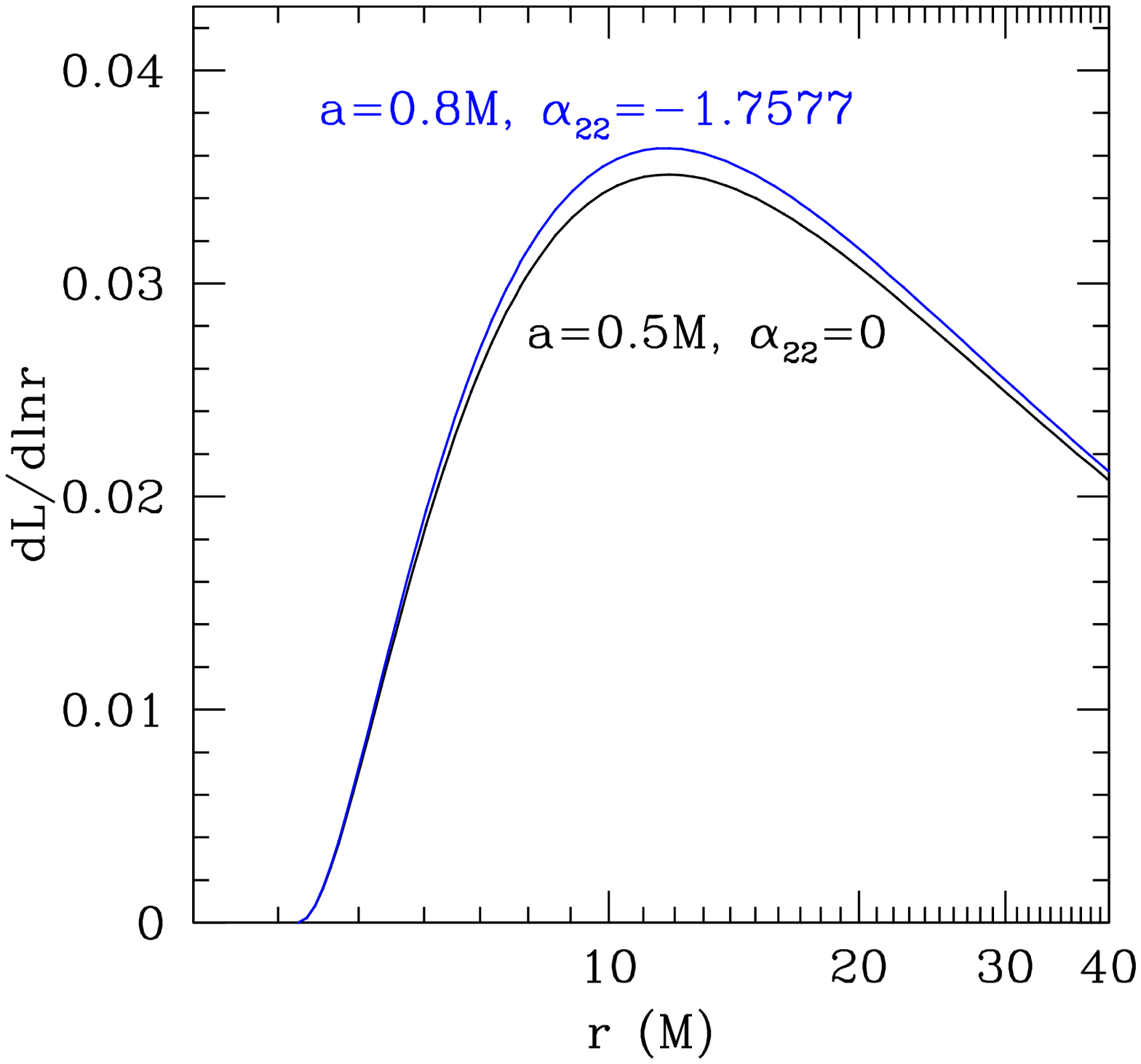,height=1.72in}
\psfig{figure=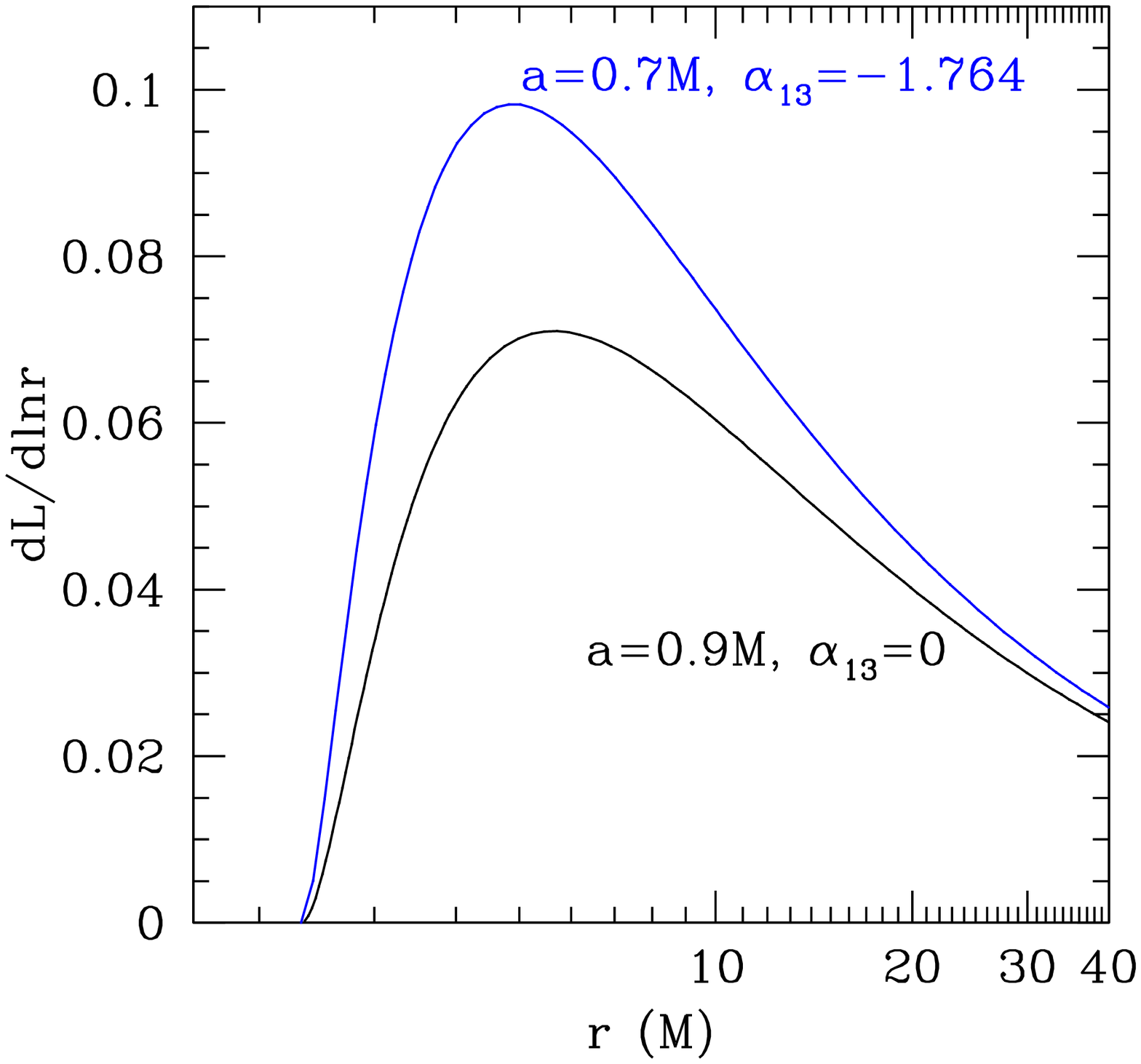,height=1.72in}
\psfig{figure=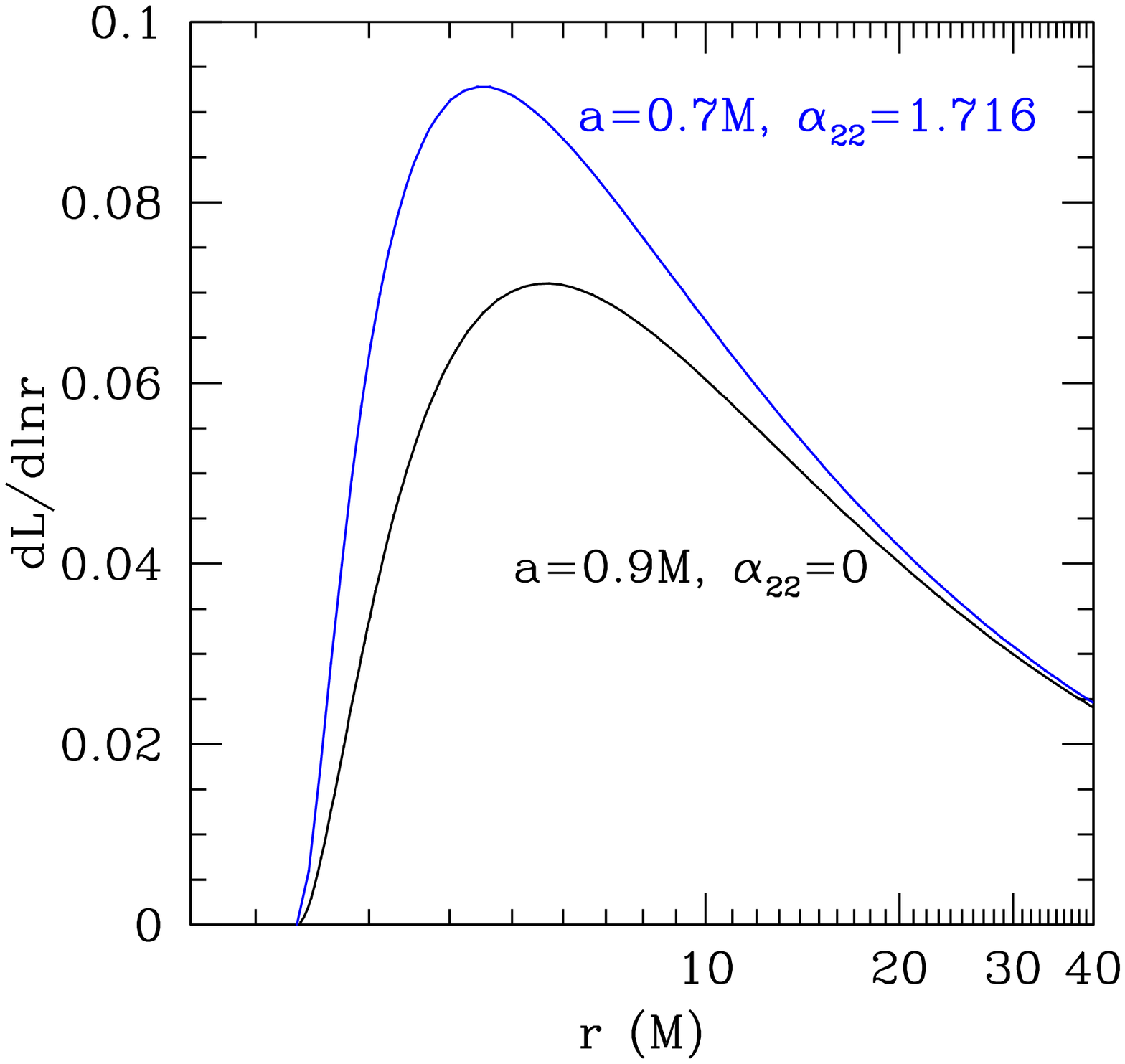,height=1.72in}
\end{center}
\caption{Radial profiles of the emitted disk luminosity (in units of $c^6/4\pi G^2$) of black holes with different values of the spin and the deviation parameters $\alpha_{13}$ and $\alpha_{22}$ such that the parameter combinations in each panel correspond to the same ISCO. The profiles are very similar for sufficiently large ISCO radii corresponding to Kerr black holes with small to intermediate spins. At high spin values, however, the emitted fluxes differ significantly.}
\label{fig:emitflux_isco}
\end{figure*}

For the accretion disk of the black hole, I assume the standard relativistic thin-disk model of Ref.~\cite{nov73}. In this model, the accretion disk lies in the equatorial plane of the black hole and is geometrically thin and optically thick. It extends from the ISCO to some outer radius $r_{\rm out}$, and the disk particles move on nearly circular equatorial orbits as they are accreted by the black hole. In my analysis, I likewise assume that there is no torque at the inner boundary of the disk which is located at the ISCO.

Using the conservation of rest mass, energy, and angular momentum of the disk particles, Ref.~\cite{PT74} derived the time-averaged thermal energy flux emitted from an accretion disk. The result of Ref.~\cite{PT74} is valid for all black holes that are stationary, axisymmetric, asymptotically flat, and reflection-symmetric across the equatorial plane. Therefore, their model applies not only to the Kerr metric, but also to the Kerr-like metric in Eq.~(\ref{eq:metric}).

The emitted energy flux is given by the expression \cite{PT74}
\be
F(r) = \frac{\dot{M}}{4\pi M^2}f_{\rm disk}(r),
\label{eq:flux_emitted}
\ee
where
$\dot{M}$ is the mass accretion rate and
\be
f_{\rm disk}(r) \equiv -\frac{d\Omega}{dr} \frac{M^2}{\sqrt{-\tilde{g}} (E-\Omega L_z)^2} \int_{r_{\rm in}}^r (E-\Omega L_z)\frac{dL_z}{d\bar{r}} d\bar{r}
\ee
is a dimensionless function of radius. In the last expression, $\tilde{g}$ is the determinant of the $(t,r,\phi)$ part of the metric. For the Kerr metric, $\sqrt{-\tilde{g}}=r$, while for the metric in Eq.~(\ref{eq:metric})
\ba
\sqrt{-\tilde{g}} &=& \frac{r}{\left| 1 + \frac{\alpha_{13}M^3}{r^3} - \frac{\alpha_{22}a^2 M^2}{r^4} + \frac{\alpha_{13}a^2 M^3}{r^5} \right| } \nn \\
&& \times \sqrt{ \frac{ \left(1 + \frac{\epsilon_3 M^3}{r^3} \right)^3 }{ 1 + \frac{\alpha_{52}M^2}{r^2} } }
\label{eq:3det}
\ea

From the energy flux in Eq.~(\ref{eq:flux_emitted}), I obtain the disk luminosity
\be
L = 2 \int_{r_{\rm in}}^r 2\pi \bar{r} F(\bar{r}) d\bar{r}.
\ee
In Fig.~\ref{fig:emitflux}, I plot the radial profile of the disk luminosity for a black hole with spin $a=0.9M$ for different values of the deviation parameters $\alpha_{13}$ and $\alpha_{22}$. The luminosity profile has a sharp peak at a radius near the ISCO and falls off quickly at larger and smaller disk radii. At the ISCO, the emitted flux is zero. As either the value of the parameter $\alpha_{13}$ decreases or the value of the parameter $\alpha_{22}$ increases, the peak of the luminosity profile increases and is located at a radius that lies closer to the ISCO. The deviation parameters $\epsilon_3$ and $\alpha_{52}$ have only a minor effect on the luminosity profile which primarily either increases or decreases its maximum slightly.

Since the ISCO in the Kerr-like metric in Eq.~$(\ref{eq:metric})$ depends on the spin and on the deviation parameters $\alpha_{13}$ and $\alpha_{22}$ (as well as to a lesser extent on the parameter $\epsilon_3$), it is important to analyze the dependence of the luminosity profile on different combinations of these parameters which correspond to the same ISCO. In Fig.~\ref{fig:emitflux_isco}, I plot several cases of luminosity profiles for Kerr black holes with low, intermediate, and high spins together with the luminosity profiles for Kerr-like black holes with different spins and nonzero values of either the parameter $\alpha_{13}$ or $\alpha_{22}$ which match the ISCO of the respective Kerr black hole.

The effect of both deviation parameters is very similar. For small to intermediate values of the spin, the luminosity profile changes only slightly even for large values of the deviation parameters, because the ISCO and the peak of the luminosity profile are located at relatively large radii. For high spin values, however, the luminosity profiles are altered significantly.

Since the disk is in thermal equilibrium, the disk flux is related to the disk temperature by the standard Stefan-Boltzmann law,
\be
F(r) = \sigma T_{\rm eff}^4(r),
\ee
with the specific intensity
\be
I_e(\nu_e) = \frac{2h\nu_e^3}{c^2}\frac{1}{f_{\rm col}^4}\frac{\Upsilon}{\exp\left(\frac{h\nu_e}{k_B T_{\rm col}}\right)-1}.
\label{eq:thermalintensity}
\ee
In these expressions, $T_{\rm eff}$ is the effective temperature, $T_{\rm col}(r)\equiv f_{\rm col}T_{\rm eff}(r)$ is the color temperature with spectral hardening factor $f_{\rm col}$, $\nu_e$ is the frequency of emission, and $h$ and $k_B$ are the Planck and Boltzmann constants, respectively. Photons emitted from the disk and observed at a large distance from the black hole experience a redshift
\be
g = \frac{E_{\rm obs}}{E_{\rm e}} = \frac{g_{\mu\nu,{\rm obs}}k^\mu_{\rm obs}u^\nu_{\rm obs}}{g_{\mu\nu,{\rm e}}k^\mu_{\rm e}u^\nu_{\rm e}},
\label{eq:redshift}
\ee
where the subscripts ``obs'' and ``e'' refer to the image plane and the accretion disk, respectively, and where $u_e$ is the Keplerian frequency of the disk particles calculated in Ref.~\cite{joh13metric}.

The factor $\Upsilon$ in Eq.~(\ref{eq:thermalintensity}) accounts for the emission type which I assume to be either isotropic or limb darkened. I use a standard limb darkening law (see Refs.~\cite{Chandra50,cunn76}) which modifies the specific intensity of the disk by the factor
\be
\Upsilon(r_e) \equiv \begin{cases} 1 & \text{isotropic radiation} \\
\frac{1}{2} + \frac{3}{4}\cos \zeta & \text{limb darkened radiation},
\end{cases}
\ee
where $r_e$ is the radius of emission. In this expression, $\zeta$ is the emission angle relative to the disk normal as measured in a local inertial frame. This angle is given by the equation
\be
\cos\zeta = \frac{g\sqrt{\eta}}{\sqrt{r_e^2 + \epsilon_3\frac{M^3}{r_e}}},
\label{eq:coszeta}
\ee
where $\eta$ is given by Eq.~(\ref{eq:eta}). I derive this expression for the emission angle in the Appendix.

%%%%%%%%%%%%%%%%%%%%%%%%%%%%%%%
\subsection{Observed spectra}
\label{sec:observedspectra}

In order to calculate the observed disk flux density, I employ the ray tracing algorithm described in the previous section. Following Ref.~\cite{li05}, I calculate the observed photon number flux density $N_{E_{\rm obs}}\equiv F_{E_{\rm obs}}/E_{\rm obs}$, which is given by the integral
\ba
N_{E_{\rm obs}} &=& \frac{1}{E_{\rm obs}} \int I_{\rm obs}(\nu_{\rm obs})d\Omega_{\rm obs} = \frac{1}{E_{\rm obs}} \int g^3 I_{\rm e}(\nu_e)d\Omega_{\rm obs} \nn \\
&=& N_0 \left( \frac{ E_{\rm obs} }{ {\rm keV} } \right)^2 \int \frac{1}{M^2} \frac{ \Upsilon r' dr'd\phi'}{\exp \left[ \frac{N_1}{g F^{1/4}} \left( \frac{ E_{\rm obs} }{ {\rm keV} } \right) \right] - 1},
\label{eq:Nobs}
\ea
where \cite{li05}
\be
N_0 \equiv 0.07205 f_{\rm col}^{-4} \left( \frac{M}{M_\odot} \right)^2 \left( \frac{D}{{\rm kpc}}\right)^{-2}{\rm keV^{-1}~cm^{-2}~s^{-1}},
\ee
\be
N_1 \equiv 0.1331 f_{\rm col}^{-1} \left( \frac{\dot{M}}{10^{18}~{\rm g~s^{-1}}} \right)^{-1/4} \left( \frac{M}{M_\odot} \right)^{1/2}
\ee
and where the Lorentz invariant $I/\nu^3$ was used in Eq.~(\ref{eq:Nobs}) to relate the observed specific intensity to the emitted specific intensity. I choose polar coordinates $(r',\phi')$ in the image plane and an outer disk radius $r_{\rm out}=10^6M$. I evaluate the integral in Eq.~(\ref{eq:Nobs}) numerically using a large number of rays with logarithmically spaced radii and equally spaced polar angles in the image plane.

For a Kerr black hole, the observed spectrum of the number flux density was discussed in great detail by, e.g., Ref.~\cite{li05}. For higher values of the spin $a$ of the black hole, the spectrum becomes harder, because the inner edge of the accretion disk extends to smaller radii, which causes the radiative efficiency and the disk temperature to increase. For larger values of the disk inclination $i$, the observed flux density at lower energies decreases, because it originates at larger disk radii, where the observed flux density is $\propto \cos i$, while the observed flux density increases at higher energies, because it originates at smaller radii, where the relativistic effects of boosting, beaming, and light bending become important. As expected, when the accretion rate increases, the observed flux density likewise increases, and the spectrum becomes harder for higher values of the spectral hardening factor. Reference~\cite{li05} also studied the effects of a nonzero torque at the inner edge of the disk and of returning radiation (cf. \cite{cunn76}). They showed that these effects can be compensated by an adjustment of the mass accretion rate and the spectral hardening factor in a disk model without returning radiation and a torque at the ISCO. The effect of limb darkening is particularly strong at high disk inclinations and leads to a lower observed flux density relative to the observed flux density when the emission is isotropic.

\begin{figure}[ht]
\begin{center}
\psfig{figure=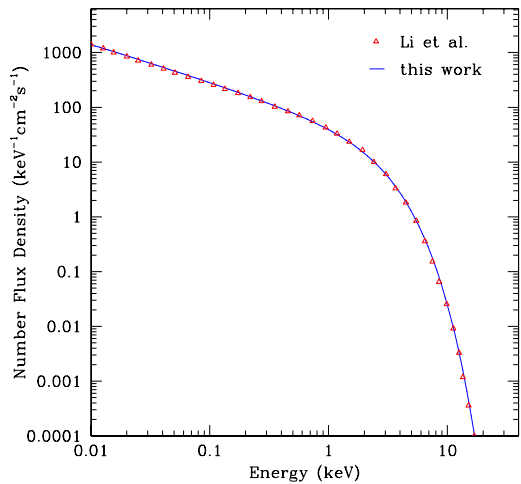,width=0.4\textwidth}
\end{center}
\caption{Observed number flux density from a geometrically thin Novikov-Thorne type accretion disk around a Kerr black hole with spin $a=0.999M$, inclination $i=30^\circ$, mass $M=10M_\odot$, distance $D=10~{\rm kpc}$, mass accretion rate $\dot{M}=10^{19}~{\rm g~s^{-1}}$, and spectral hardening factor $f_{\rm col}=1$. The red triangles denote the number flux density reported by Ref.~\cite{li05}, while the solid blue line is the number flux density calculated with the algorithm described here.}
\label{fig:crosscheck}
\end{figure}

I verified these spectral properties for Kerr black holes using the algorithm described in Sec.~\ref{sec:code}. In Fig.~\ref{fig:crosscheck}, I compare one spectrum generated by this algorithm with a spectrum from Ref.~\cite{li05} with the same parameters and find excellent agreement. Note, however, that my algorithm does not incorporate the effects of returning radiation or a nonzero torque at the ISCO. While these effects can be absorbed into modified values of the mass accretion rate and the spectral hardening factor in the case of a Kerr black hole as shown by Ref.~\cite{li05}, it is not obvious that this adjustment works equally well in the extended theory where the black hole is described by the metric in Eq.~(\ref{eq:metric}), because the deviation parameters may have different effects on the light bending and the orbital velocity of the disk plasma at the ISCO. Nonetheless, one would expect that such an adjustment of the mass accretion rate and the spectral hardening factor can compensate for these effects at least approximately, especially when the deviations from the Kerr metric are small.

\begin{figure}[ht]
\begin{center}
\psfig{figure=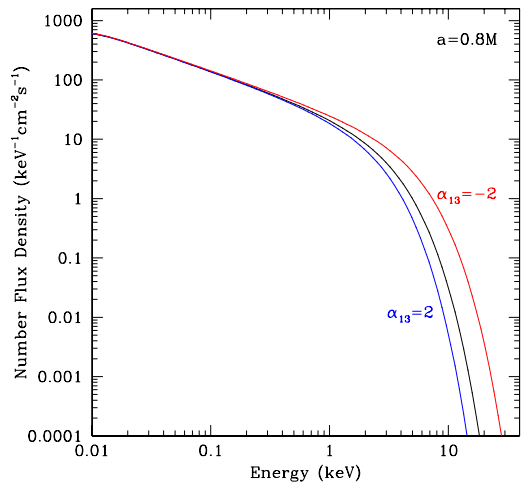,width=0.4\textwidth}
\psfig{figure=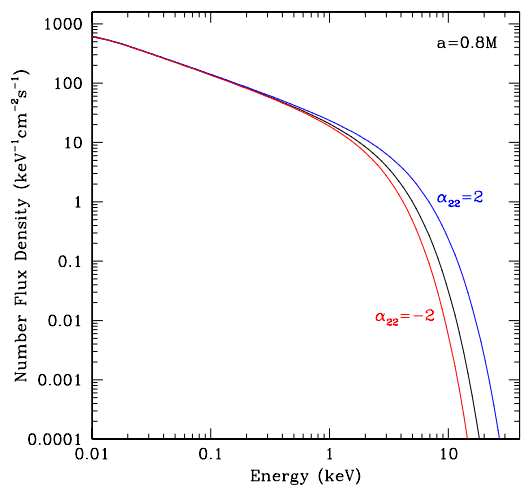,width=0.4\textwidth}
\end{center}
\caption{Observed thermal spectra from a geometrically thin accretion disk around black holes with spin $a=0.8M$, inclination $i=30^\circ$, mass $M=10M_\odot$, distance $D=10~{\rm kpc}$, mass accretion rate $\dot{M}=10^{19}~{\rm g~s^{-1}}$, and spectral hardening factor $f_{\rm col}=1.7$ for different values of the deviation parameters $\alpha_{13}$ (top) and $\alpha_{22}$ (bottom). The spectra become harder for decreasing values of the parameter $\alpha_{13}$ and increasing values of the parameter $\alpha_{22}$. For all spectra, an isotropic disk emission is assumed.}
\label{fig:spectra08}
\end{figure}

\begin{figure*}[ht]
\begin{center}
\psfig{figure=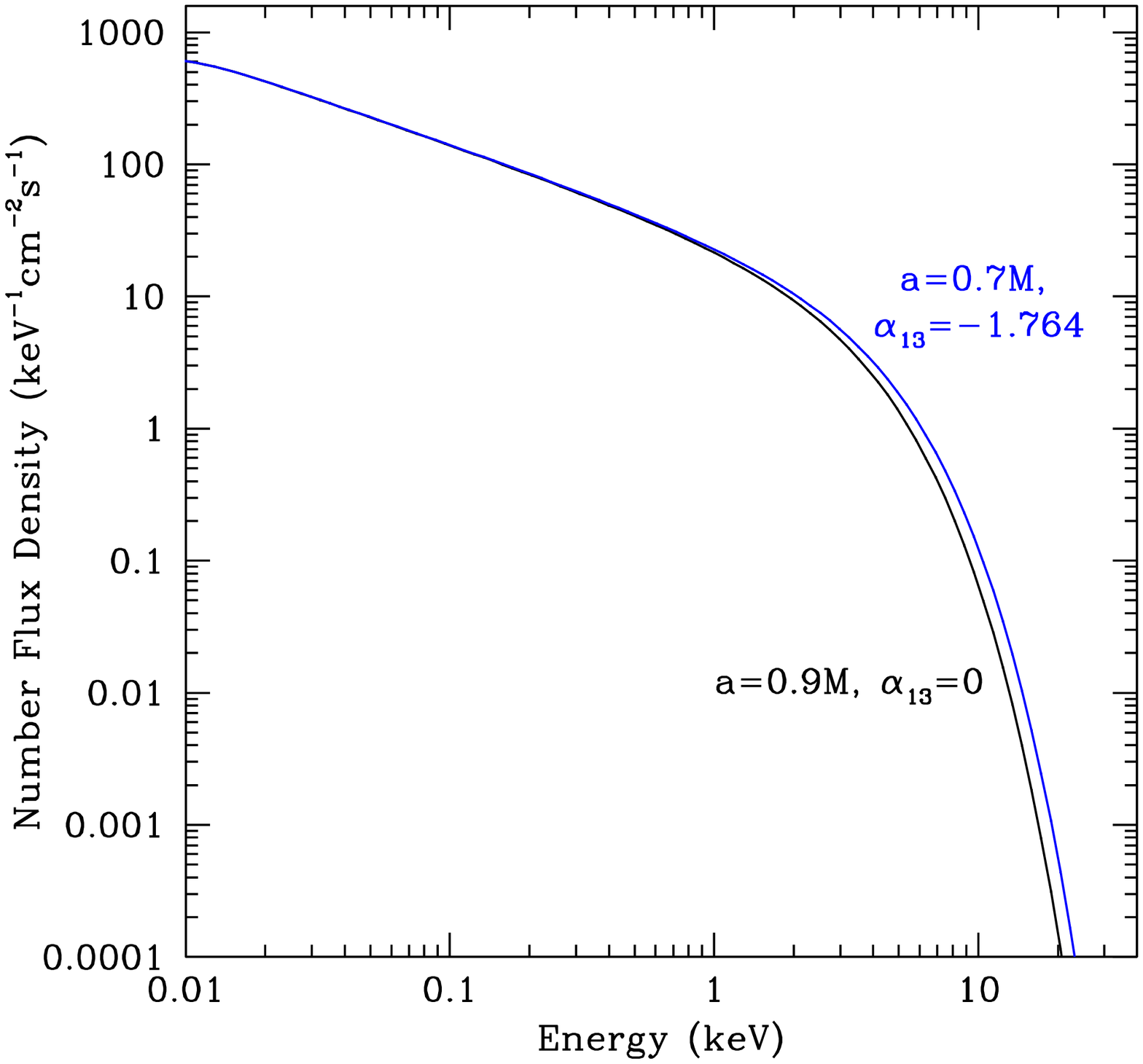,width=0.32\textwidth}
\psfig{figure=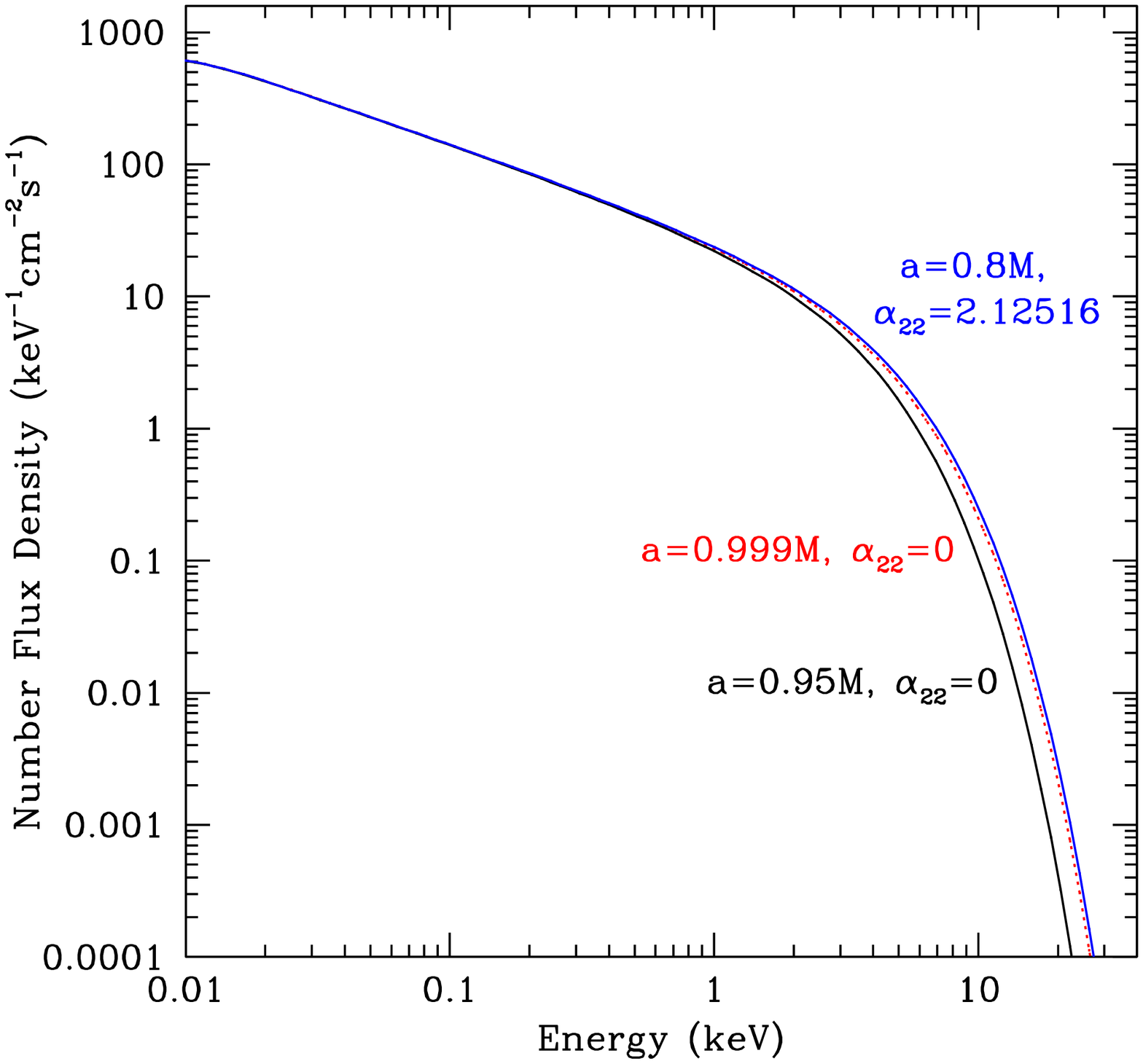,width=0.32\textwidth}
\psfig{figure=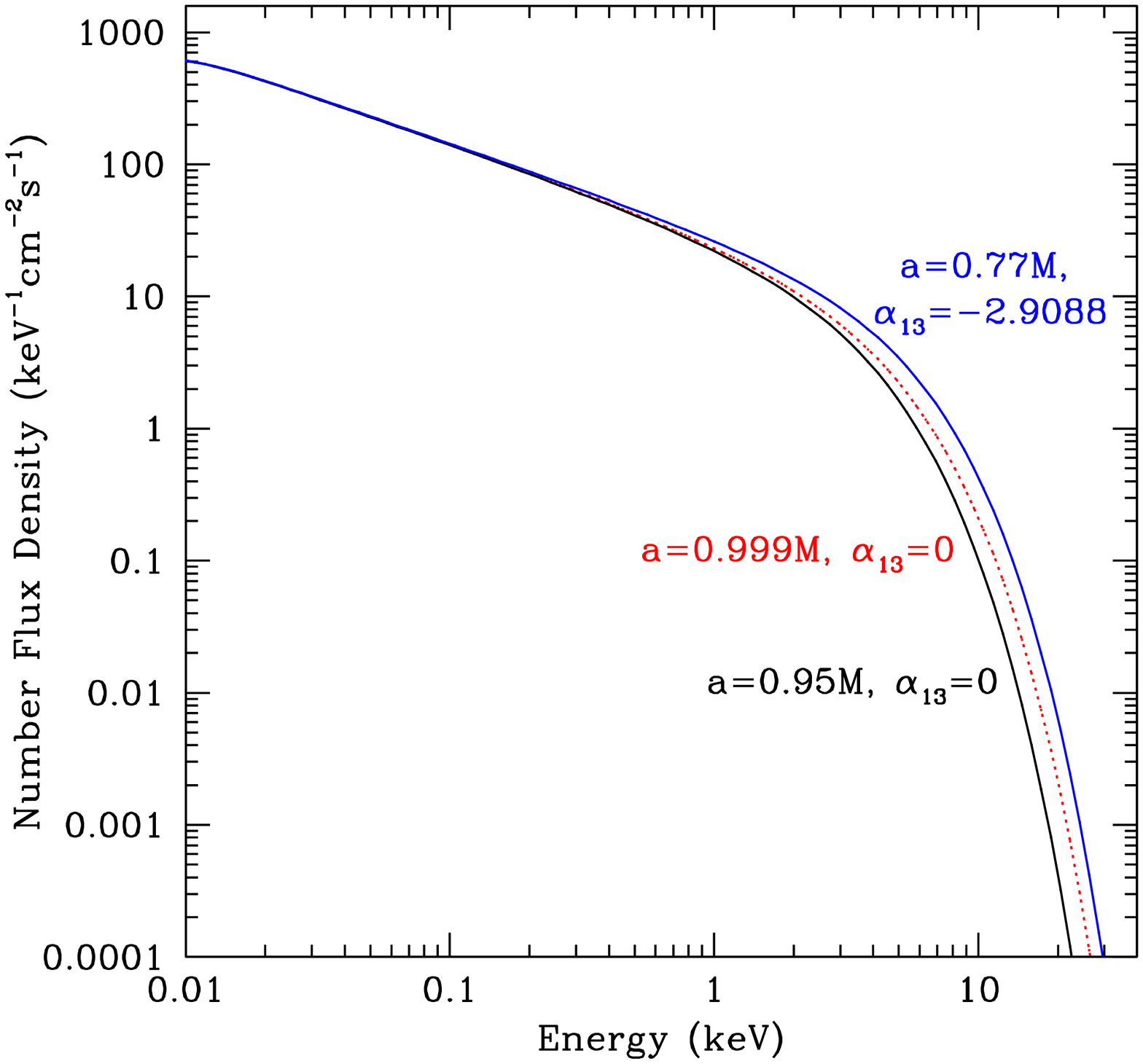,width=0.32\textwidth}
\end{center}
\caption{Observed spectra of the number flux density from a geometrically thin accretion disk around black holes with different values of the spin and deviation parameters $\alpha_{13}$ and $\alpha_{22}$ such that the ISCO coincides for the parameter combinations shown in each panel. The other parameters are the inclination $i=30^\circ$, the mass $M=10M_\odot$, the distance $D=10~{\rm kpc}$, the mass accretion rate $\dot{M}=10^{19}~{\rm g~s^{-1}}$, and the spectral hardening factor $f_{\rm col}=1.7$. The spectra differ significantly at the high-energy end, but some of the spectra of the Kerr-like black holes can still be confused with the spectra of a Kerr black hole with a different ISCO and spin. In the left panel, the spectrum of the Kerr-like black hole is very similar to the spectrum of a Kerr black hole with a spin $a\approx0.96M$ (not shown). In the center panel, the spectrum of the Kerr-like black hole can barely be mimicked by the spectrum of a Kerr black hole with a spin $a=0.999M$ (red dotted curve). In the right panel, however, the spectrum of the Kerr-like black hole cannot originate from a Kerr black hole even with a spin $a=0.999M$ (red dotted curve), because it does not extend to such high energies.}
\label{fig:spectra_isco}
\end{figure*}

The effect of the deviations from the Kerr metric on the observed spectra depend on the particular parameter. The spectra are strongly affected by both the parameters $\alpha_{13}$ and $\alpha_{22}$ in a manner that is similar to the effect of changing the spin of the black hole. In Fig.~\ref{fig:spectra08}, I plot the observed spectrum of the number flux density for a black hole with mass $M=10M_\odot$ and spin $a=0.8M$ for different values of the deviation parameters $\alpha_{13}$ and $\alpha_{22}$. When one deviation parameter is varied, the other one is set to zero. The other system parameters are the inclination $i=30^\circ$, the distance $D=10~{\rm kpc}$, the mass accretion rate $\dot{M}=10^{19}~{\rm g~s^{-1}}$, and the spectra hardening factor $f_{\rm col}=1.7$, and I assume an isotropic disk emission. For decreasing values of the parameter $\alpha_{13}$ and increasing values of the parameter $\alpha_{22}$, the number flux density extends to higher energies, while the low-energy part of these spectra is practically unaffected. Since the deviation parameters predominantly affect the inner part of the disk, this is as expected. Changing the disk inclination if one or more of the deviation parameters are nonzero has qualitatively the same effect as in the case of a Kerr black hole.

Since the emitted flux depends only marginally on the parameters $\epsilon_3$ and $\alpha_{52}$, the effect of these parameters on the observed spectra is very small, at least in the case of isotropic emission. Both parameters primarily affect the peak flux density of the emitted radiation and the overall normalization of the observed flux density through the determinant of the metric in Eq.~(\ref{eq:flux_emitted}), while nonzero values of the parameter $\epsilon_3$ also cause a slight shift of the location of the ISCO (see Fig.~6 in Ref.~\cite{joh13metric}). Since the normalization of the flux density is used in practice to infer the mass accretion rate or, equivalently, the disk luminosity in units of the Eddington luminosity (see, e.g., Ref.~\cite{mccl06}), the parameters $\epsilon_3$ and $\alpha_{52}$ cannot be obtained from the flux density normalization. Conversely, since their effect on the normalization is negligible, the mass accretion rate can be inferred robustly even if these two parameters are nonzero.

In the case of limb darkened emission, the deviation parameters $\epsilon_3$, $\alpha_{13}$, and $\alpha_{22}$ have an effect on the observed spectra that is comparable in magnitude and analogous to the effect of limb darkening for Kerr black holes (cf. \cite{li05}). As before, the parameter $\alpha_{52}$ has only a very minor effect on the observed spectra. Consequently, the observed flux density depends strongly on the deviation parameters $\alpha_{13}$ and $\alpha_{22}$ in the case of isotropic emission, while the observed flux density depends strongly on the deviation parameters $\epsilon_3$, $\alpha_{13}$, and $\alpha_{22}$ in the case of limb darkened emission. In all cases, these parameters affect the observed spectra in a manner that is different from the effect of changing the mass accretion rate, which can, therefore, be measured independently of these parameters from the normalization of the flux density.

\begin{figure*}[ht]
\begin{center}
\psfig{figure=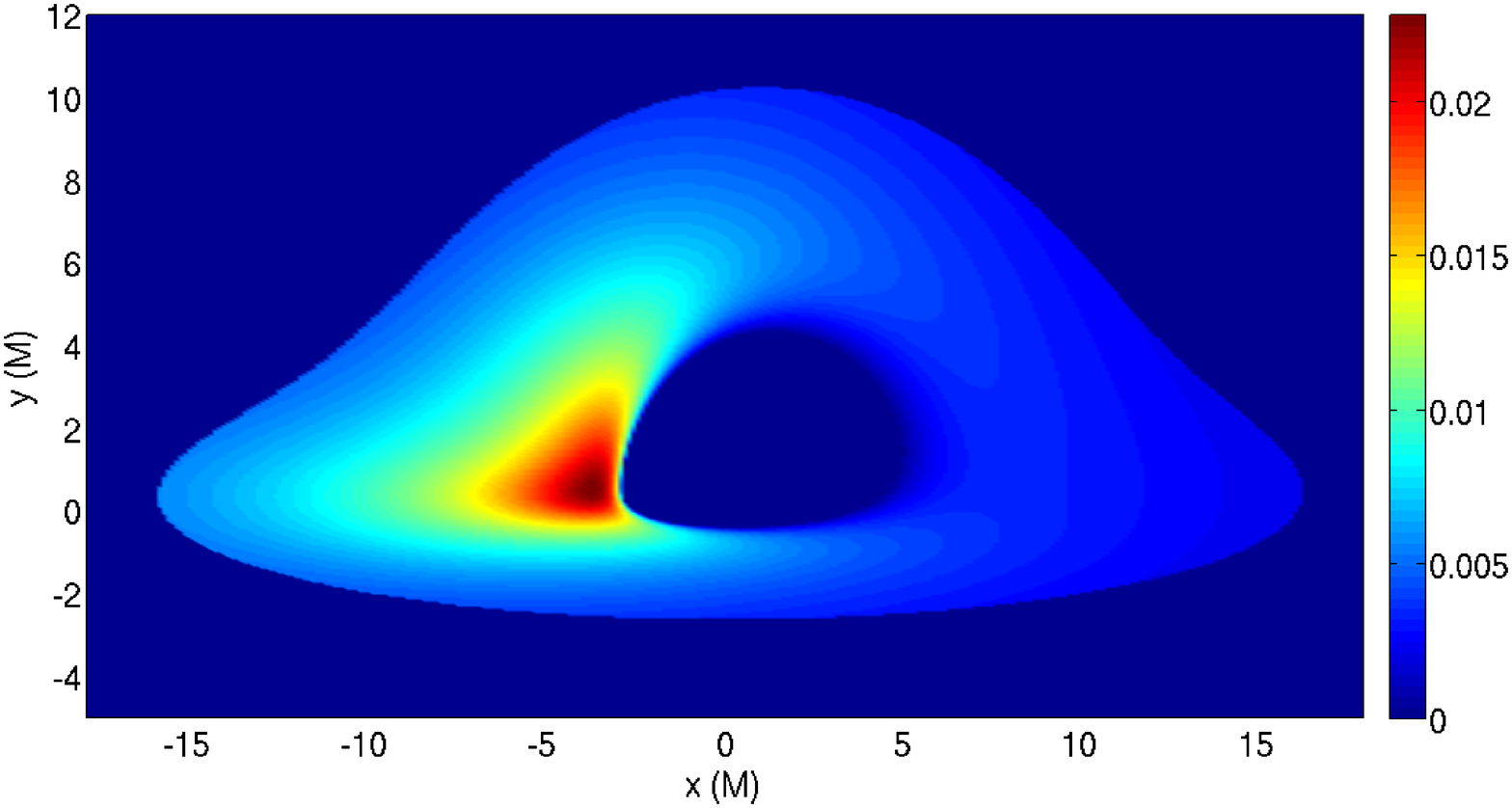,height=1.62in}
\psfig{figure=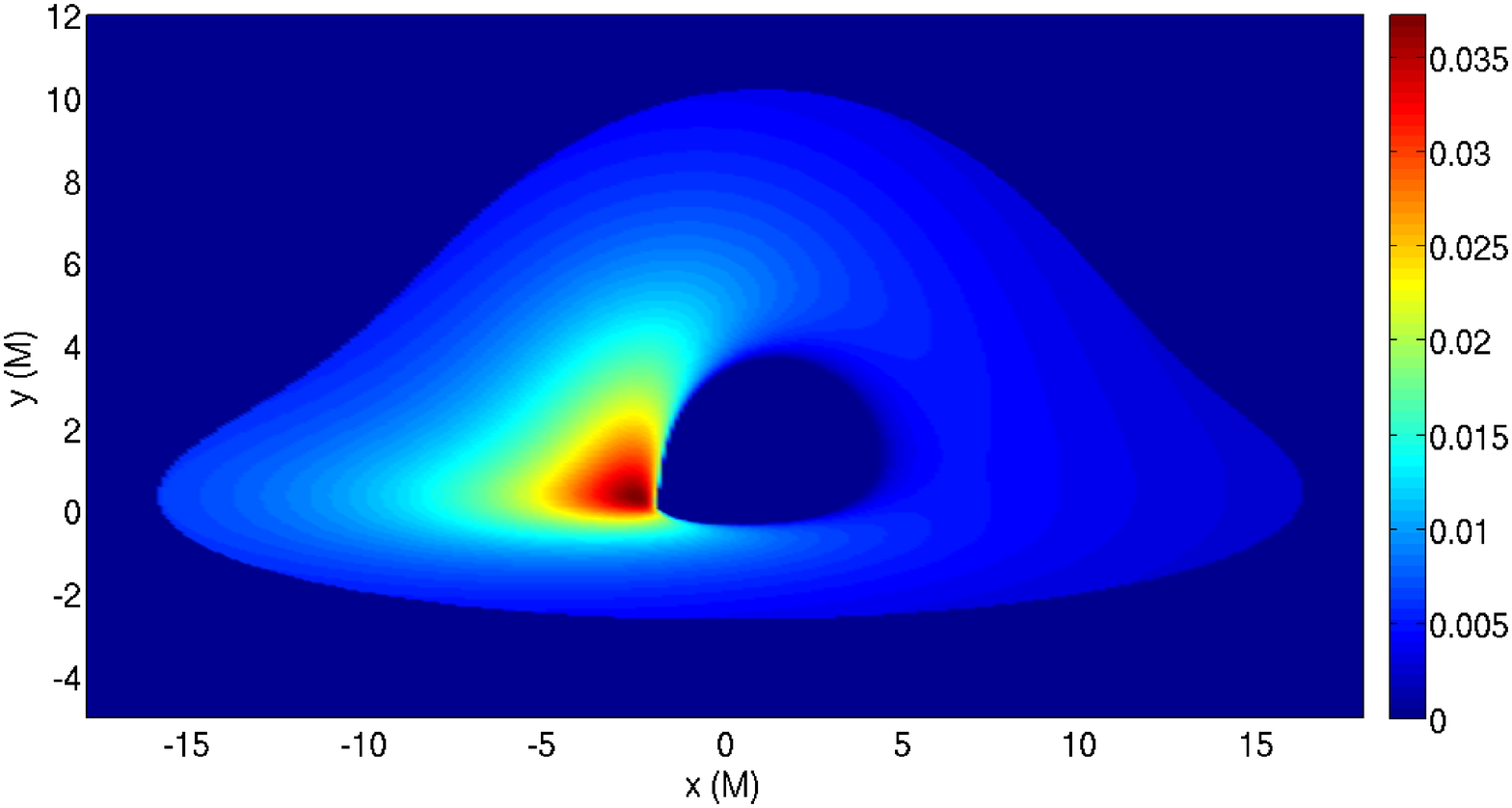,height=1.62in}
\end{center}
\caption{Direct images of the inner parts of geometrically thin accretion disks around (left) a Kerr black hole with spin $a=0.95M$ and (right) a Kerr-like black hole with the same spin and a value of the deviation parameter $\alpha_{13}=-1$ for fixed values of the inclination $i=80^\circ$, mass $M=10M_\odot$, distance $D=10~{\rm kpc}$, mass accretion rate $\dot{M}=10^{19}~{\rm g~s^{-1}}$, and spectral hardening factor $f_{\rm col}=1.7$. Both panels show the observed number flux density at $1~{\rm keV}$ in units of ${\rm keV^{-1}~cm^{-2}~s^{-1}}$. The highest emission originates from a strongly localized region near the ISCO on the side of the black hole that is approaching the observer where Doppler boosting and beaming are particularly high. This region shifts toward the black hole and emits a significantly higher flux for negative values of the parameter $\alpha_{13}$.}
\label{fig:diskimage}
\end{figure*}

Since the ISCO depends on the spin and the deviation parameters $\epsilon_3$, $\alpha_{13}$, and $\alpha_{22}$, the question of whether the spin and the deviation parameters can be measured independently has to be analyzed more carefully. Since the emitted flux of a Kerr-like black hole only differs significantly from the emitted flux of a Kerr black hole with the same ISCO radius if the latter black hole is spinning rapidly as can be seen from Fig.~\ref{fig:emitflux_isco}, one would expect that the observed flux density is very similar for Kerr and Kerr-like black holes with the same ISCO as long as the Kerr black hole has a small to intermediate spin. This is indeed the case and the corresponding observed spectra are practically indistinguishable.

For high values of the spin, however, the observed spectra differ significantly. In Fig.~\ref{fig:spectra_isco}, I illustrate this for particular choices of the deviation parameters $\alpha_{13}$ and $\alpha_{22}$. Fig.~\ref{fig:spectra_isco} shows the observed number flux density for Kerr black holes with spins $a=0.9M$ and $a=0.95M$ and for Kerr-like black holes with smaller spins but a nonzero values of the parameters $\alpha_{13}$ or $\alpha_{22}$ chosen such that the ISCO coincides with the ISCO of the respective Kerr black hole. The observed spectra differ strongly at the high-energy end. Nonetheless, spectra of Kerr-like black holes with nonzero values of the deviation parameters are still very similar to the spectra of Kerr black holes with a different spin and, therefore, a different ISCO. However, if the spin of the Kerr black hole is near maximal, spectra of Kerr-like black holes can extend to such high energies that they cannot originate from a maximally spinning Kerr black hole.

These results imply that the deviation parameters $\alpha_{13}$ and $\alpha_{22}$ can be measured independently of the spin if the radius of the ISCO is $\approx M$ and the magnitude of the deviation is of order unity. This also means that the location of the ISCO can be measured robustly as long as the ISCO radius is not too close to the black hole corresponding to Kerr black holes with small to intermediate values of the spin. Since the limb darkening does not affect the high-energy end of the spectrum, nonzero values of the parameter $\epsilon_3$ can always be closely mimicked by a Kerr black hole with a different spin.

In Fig.~\ref{fig:diskimage}, I illustrate the appearance of accretion disks for black holes with mass $M=10M_\odot$, spin $a=0.95M$, and different values of the deviation parameter $\alpha_{13}$ as ``seen'' by a distant observer at an energy of $1~{\rm keV}$. The other parameters are held fixed with values of the inclination $i=80^\circ$, distance $D=10~{\rm kpc}$, mass accretion rate $\dot{M}=10^{19}~{\rm g~s^{-1}}$, and spectral hardening factor $f_{\rm col}=1.7$. Here I focus only on the innermost part of the disk with a radial extent stretching from the ISCO to $r=15M$. This part of the disk experiences the most dramatic effects if the deviation parameters are nonzero. The observed flux density is strongly peaked on the side of the disk that approaches the observer due to enhanced Doppler boosting and beaming. Decreasing the value of the parameter $\alpha_{13}$ increases the peak flux drastically. Likewise, the ISCO radius decreases and the location of the peak flux shifts to a smaller radius.

Because of their minute angular sizes, however, such images of accretions disks are unobservable for stellar-mass black holes. While the microquasar GRS~1915+105 has a huge accretion disk with an estimated radius of $\sim10^{12}~{\rm cm}$ \cite{done04} that would be large enough to be imaged with a very-long baseline interferometer (see Ref.~\cite{fish13}), its innermost region will not be resolved. In the case of supermassive black holes such as the one at the Galactic center, the observed spectra are much better described by radiatively inefficient accretion flow models (e.g., \cite{RIAF}).

%%%%%%%%%%%%%%%%%%%%%%%%%%%%%%%%%%%%%%%
\section{RELATIVISTICALLY BROADENED IRON LINES}
\label{sec:ironlines}

In this section, I simulate relativistically broadened iron lines. As in Refs.~\cite{PJ12,jp13}, I assume that the disk plasma moves on circular equatorial orbits at the local Keplerian velocity. I further assume that the emission from the disk is either isotropic or limb darkened as well as monochromatic with a rest frame energy $E_0$ (e.g., $E_0\approx6.4~{\rm keV}$ for the iron K$\alpha$ line) and has an emissivity $\epsilon(r)\propto r^{-\alpha}$, where $\alpha$ is the emissivity index.

The observed specific flux is then given by the expression
\be
F_E = \frac{1}{D^2} \int dx' \int dy' I(x',y') \delta[E_e-E_0 g(x',y')],
\ee
which I evaluate using the algorithm described in Refs.~\cite{PJ12,jp13} with the metric in Eq.~(\ref{eq:metric}) as the underlying spacetime and the changes outlined in Section~\ref{sec:code}.

\begin{figure}[ht]
\begin{center}
\psfig{figure=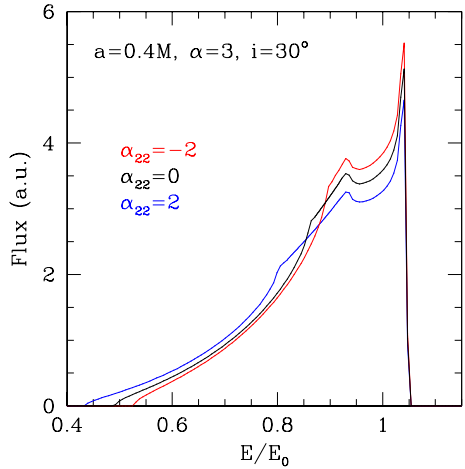,width=0.4\textwidth}
\psfig{figure=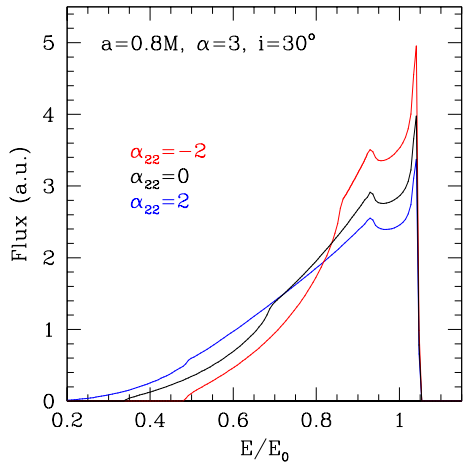,width=0.4\textwidth}
\end{center}
\caption{Iron line profiles for black holes with spins (top) $a=0.4M$ and (bottom) $a=0.8M$ with an outer disk radius $r_{\rm out}=100M$, a disk inclination $i=30^\circ$, and an emissivity index $\alpha=3$ for several values of the deviation parameter $\alpha_{22}$. The energy $E$ is measured in units of the energy of emission $E_0$. The line profiles are altered primarily at high energies and in their extent toward low energies.}
\label{fig:lines30}
\end{figure}

The effect of the deviation parameters on observed iron line profiles is similar to their effects on the thermal disk spectra described in Sec.~\ref{sec:observedspectra}. In Fig.~\ref{fig:lines30}, I plot iron line profiles for black holes with spins $a=0.4M$ and $a=0.8M$ for several values of the deviation parameter $\alpha_{22}$. The disk inclination is $i=30^\circ$, the outer disk radius is $r_{\rm out}=100M$, the emissivity index is $\alpha=3$, and the emission is isotropic. The energy $E$ is displayed in units of the energy of emission $E_0$. For decreasing values of the parameter $\alpha_{22}$, the fluxes of the ``blue'' and ``red'' peaks increase and the ``red tail'' of the line profile is shortened. The first effect is primarily caused by the orbital velocity of the accretion flow, while the second effect is determined by the location of the ISCO and the photons that are emitted near the ISCO, which experience a strong gravitational redshift.

The effect of the parameter $\alpha_{13}$ is very similar with the difference that the corresponding modification of the line profile as mentioned above occurs for increasing values of the parameter $\alpha_{13}$ instead of decreasing values. Similar to the thermal disk spectra, the iron line profiles depend only marginally on the parameters $\epsilon_3$ and $\alpha_{52}$ if the emission is isotropic.

\begin{figure}[ht]
\begin{center}
\psfig{figure=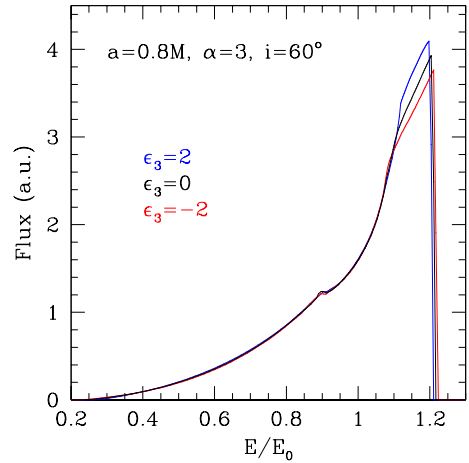,width=0.4\textwidth}
\end{center}
\caption{Iron line profiles for black holes with spin $a=0.8M$, an outer disk radius $r_{\rm out}=100M$, a disk inclination \mbox{$i=60^\circ$}, and an emissivity index $\alpha=3$ for several values of the parameter $\epsilon_3$. In all cases, the disk emission is limb darkened and the line energy $E$ is measured in units of the emission energy $E_0$. For increasing values of the parameter $\epsilon_3$, the peak flux increases.}
\label{fig:eplines}
\end{figure}

\begin{figure*}[ht]
\begin{center}
\psfig{figure=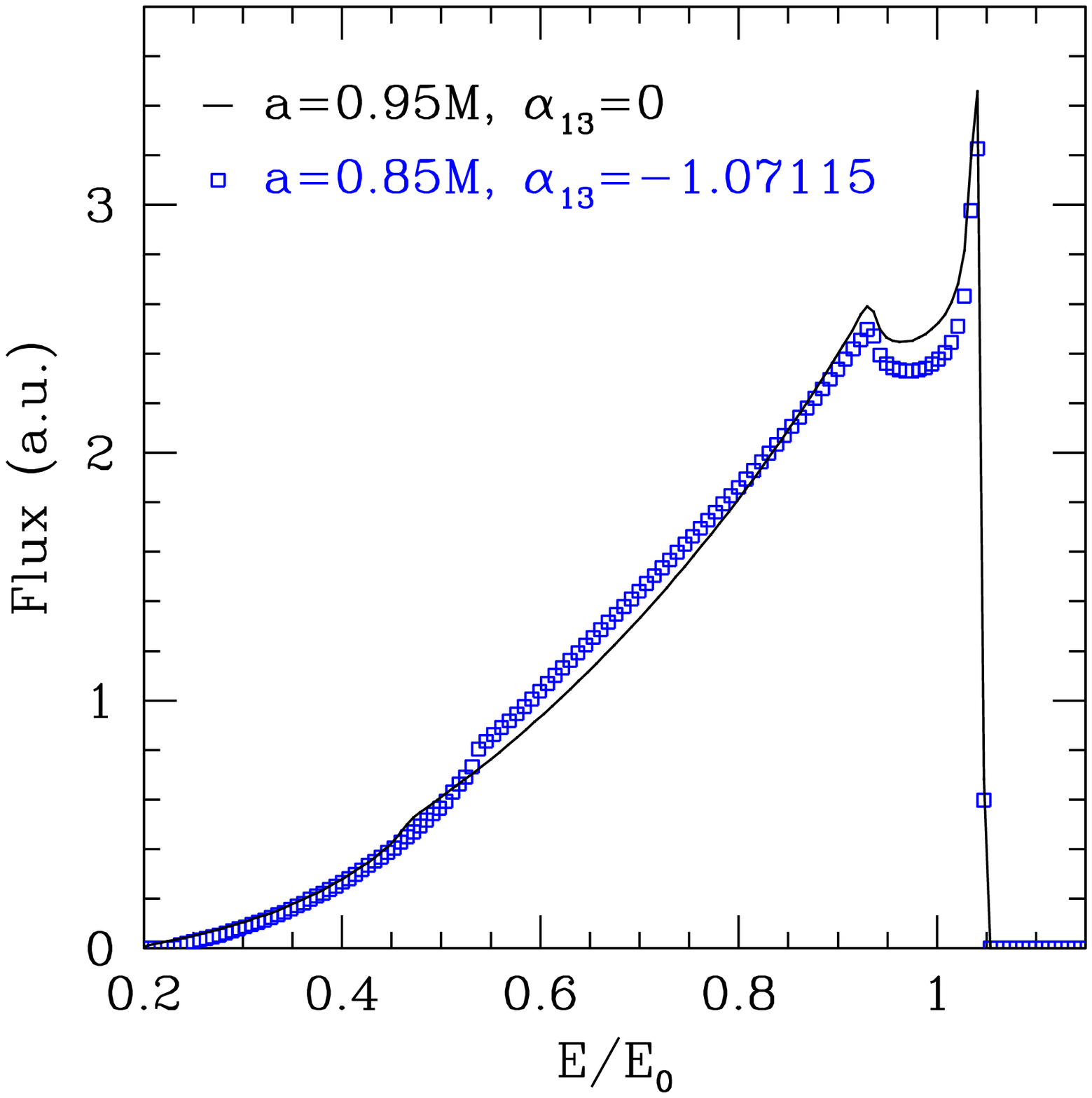,width=0.32\textwidth}
\psfig{figure=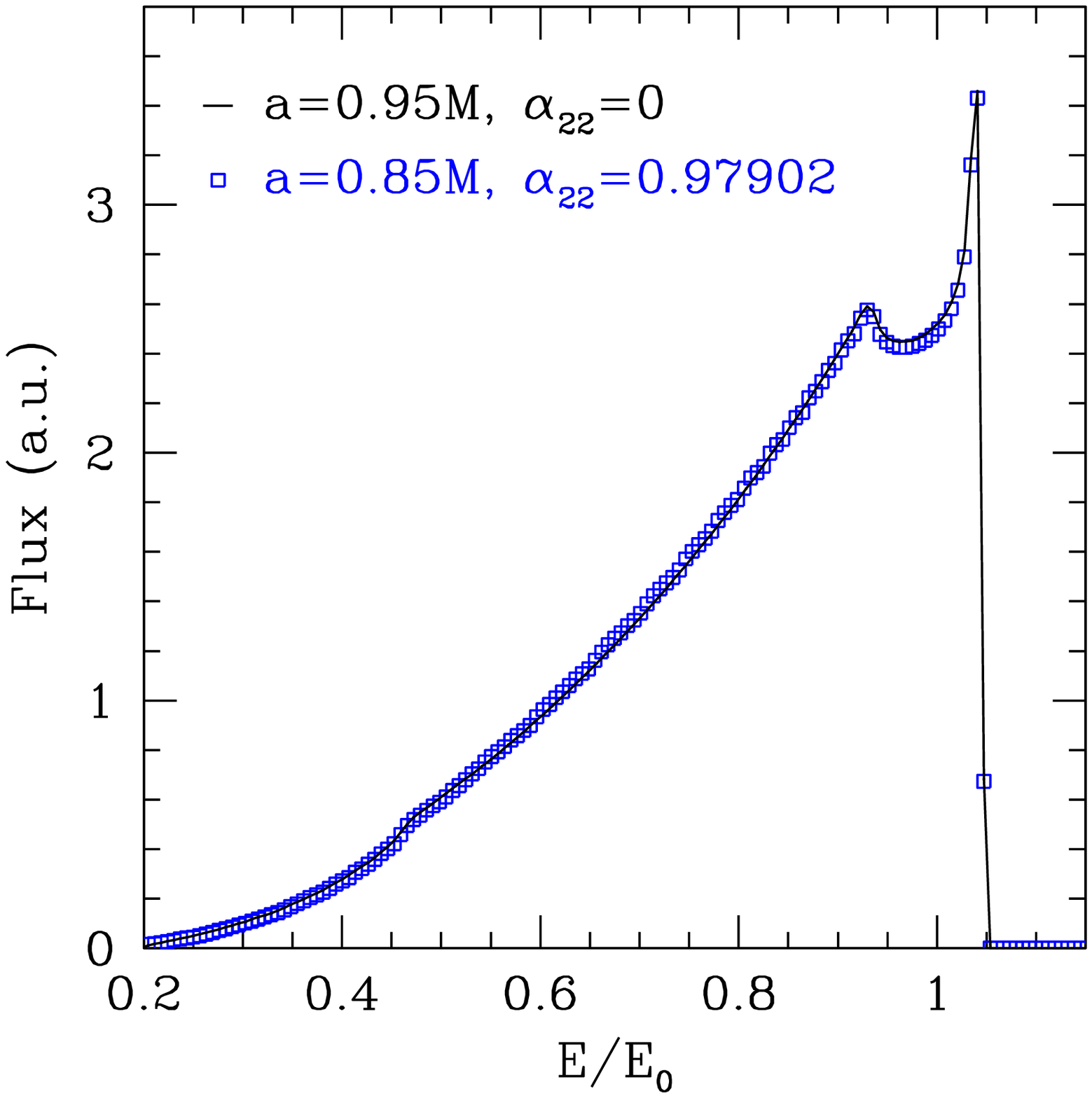,width=0.32\textwidth}
\psfig{figure=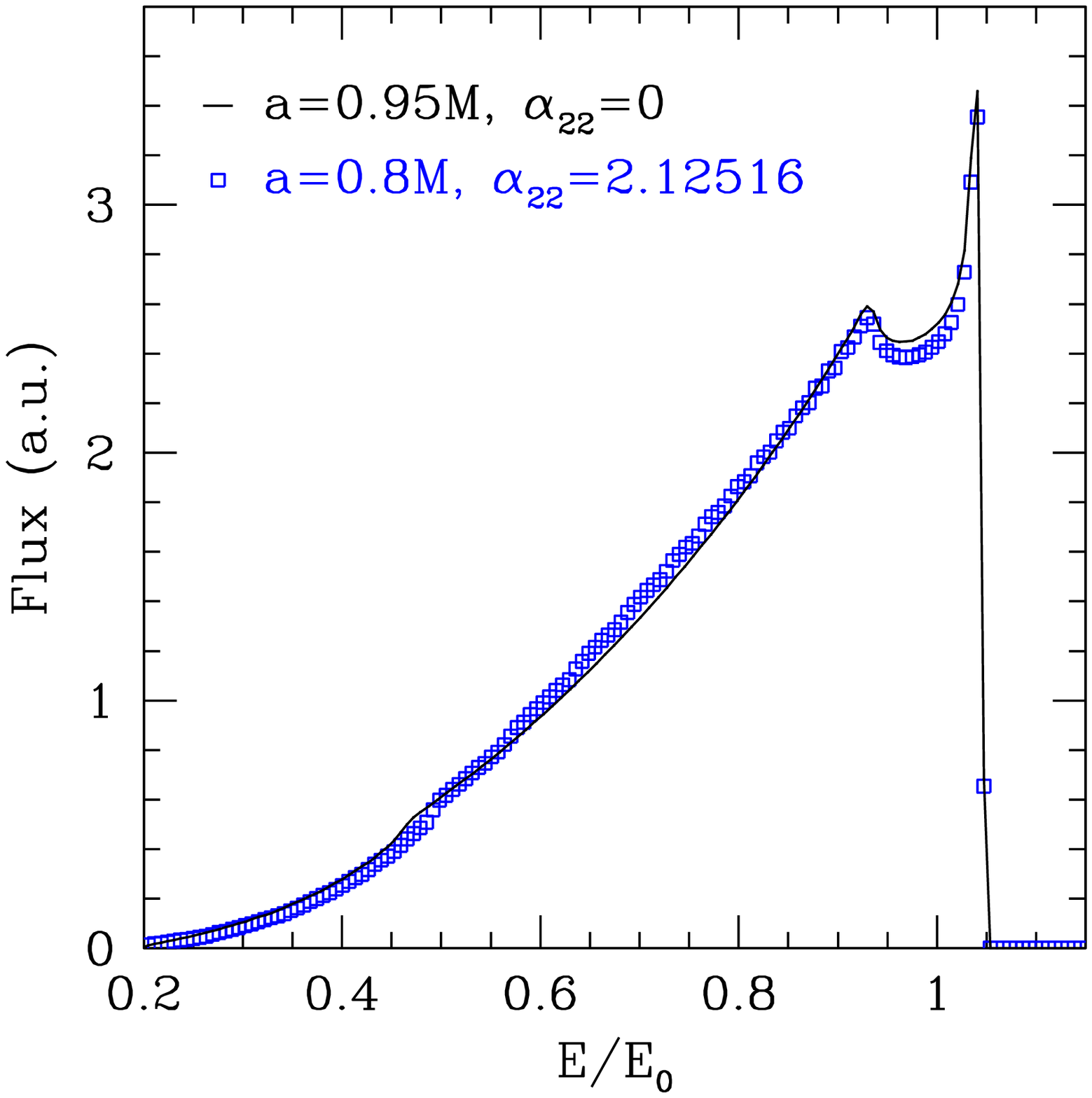,width=0.32\textwidth}
\end{center}
\caption{Iron line profiles for different values of the spin and the deviation parameters $\alpha_{13}$ (top) and $\alpha_{22}$ (bottom) such that for both sets of parameters in each panel the ISCO coincides. The other system parameters are held fixed with values $r_{\rm out}=100M$, $i=30^\circ$, and $\alpha=3$. For Kerr black holes with low to intermediate values of the spin, the line profiles are very similar to the profiles for Kerr-like black holes which have the same ISCO. For Kerr black holes with high spins, however, the line profiles are different from the profiles which correspond to the same ISCO radius with nonzero values of the deviation parameters $\alpha_{13}$ and $\alpha_{22}$ if the deviation is sufficiently large.}
\label{fig:iscolines}
\end{figure*}

At higher disk inclinations and for nonzero values of the parameters $\alpha_{13}$ or $\alpha_{22}$, the blue peak is slightly altered, while the red peak is affected only marginally. Reference~\cite{jp13} observed a similar effect in the case of iron line profiles described by their Kerr-like metric which depends on only one deviation parameter \cite{JPmetric} and identified the flux ratio of the two peaks as a potential observable of a violation of the no-hair theorem. In the metric I use in this paper, this effect still prevails. However, for line profiles simulated in both spacetimes (as well as the Kerr metric) the red peak is often submerged into the line profile and may be difficult to identify in practice. Similar changes of the peak flux can also be achieved by different values of the emissivity index and the outer disk radius as is well known for Kerr black holes. These parameters have to be determined from a spectral fit of the entire line profile.

If the emission is limb darkened, changing the parameters $\alpha_{13}$ and $\alpha_{22}$ has a similar effect on the line profiles as for Kerr black holes and the line profiles appear slightly narrower with an increased peak flux and, at lower inclinations, a decreased flux at lower energies. For nonzero values of the parameters $\epsilon_3$ and $\alpha_{52}$, the line profiles remain nearly unaffected at lower inclinations. At higher inclinations, however, the peak flux is modified and increases for positive values of the parameters $\epsilon_3$ and $\alpha_{52}$ and decreases for negative values of the parameters $\epsilon_3$ and $\alpha_{52}$, which I illustrate in Fig.~\ref{fig:eplines}. Note that this modification is slightly different from the one caused by nonzero values of the deviation parameters $\alpha_{13}$ and $\alpha_{22}$ for either isotropic or limb darkened emission.

As in the case of Kerr black holes, for iron lines that originate from the Kerr-like compact objects that are described by the metric proposed by Ref.~\cite{JPmetric}, Ref.~\cite{jp13} found that at least for small to intermediate disk inclinations the inclination angle can be robustly determined from the location of the ``blue edge'' of the line, which depends only very little on the other system parameters including the deviation from the Kerr metric. For iron line profiles simulated in a background described by the metric in Eq.~(\ref{eq:metric}), this is likewise the case.

References~\cite{PJ12,jp13} also found that for arbitrary spins over the entire spin range iron lines of Kerr-like compact objects described by either their metric or the quasi-Kerr metric are strongly correlated with the iron lines of Kerr black holes if the spin and deviation parameter are chosen such that the ISCO (or, more generally, the inner edge of the accretion disk \cite{joh13edges}) in both cases coincides. In the new Kerr-like metric, for nonzero values of the parameters $\alpha_{13}$ and $\alpha_{22}$, such profiles are likewise practically indistinguishable if the Kerr black hole is spinning slowly or moderately. For Kerr black holes with high spins, however, this strong correlation between the spin and the deviation parameters $\alpha_{13}$ and $\alpha_{22}$ does not persist and the line profiles can be significantly different from each other if the deviation is large enough. I illustrate this in Fig.~\ref{fig:iscolines}. In practice, the emissivity index is a function of the radius and can reach much larger values ($\sim7$ or higher) near the ISCO. In this case, iron line profiles differ even at small values of the deviation parameters $\alpha_{13}$ and $\alpha_{22}$, because the innermost part of the accretion disk where the deviations are most important due to the strong relativistic effects contribute significantly more to the total observed line flux.

In their initial study of relativistically broadened iron lines in a non-Kerr background, Ref.~\cite{PJ12} suggested that this strong correlation should indeed be broken for rapidly spinning black holes, because the accretion disk in this case extends almost all the way down to the event horizon, where the effect of a deviation from the Kerr metric is most apparent. While this did not occur in the metric of Ref.~\cite{JPmetric}, the suggestion by Ref.~\cite{PJ12} is indeed verified for deviations described by the parameters $\alpha_{13}$ and $\alpha_{22}$ in the metric given by Eq.~(\ref{eq:metric}). Although it is difficult to compare these two metrics directly, because they are of two distinct forms, the reduced correlation seems to be caused by the different dependence of the respective metric elements on the deviation parameters. The two deviation parameters $\alpha_{13}$ and $\alpha_{22}$ in the metric used in this paper affect only its $(t,t)$, $(t,\phi)$, and $(\phi,\phi)$ components. In contrast, the metric of Ref.~\cite{JPmetric} depends on only one deviation parameter that affects all nonzero metric components apart from the $(\theta,\theta)$ component.

In addition, contrary to the case of the thermal disk spectra analyzed in Section~\ref{sec:observedspectra}, iron lines in this part of the parameter space cannot be mimicked by the profiles of black holes with a different spin, because this would manifest in an altered tail of the line at low energies due to the effect of the gravitational redshift. This illustrates an important difference between both methods: At high spins, thermal disk spectra already differ when the deviation parameters $\alpha_{13}$ or $\alpha_{22}$ are small, while iron line profiles differ only if the deviations or the emissivity index are sufficiently large. On the other hand, for sets of (high) spin values and deviation parameters $\alpha_{13}$ and $\alpha_{22}$ with different ISCO radii, thermal spectra can still be very similar, while iron lines cannot be confused, at least in principle.

%%%%%%%%%%%%%%%%%%%%%%%%%%%%%%%%%%%%%%%
\section{QUASIPERIODIC VARIABILITY}
\label{sec:qpo}

\begin{figure*}[ht]
\begin{center}
\psfig{figure=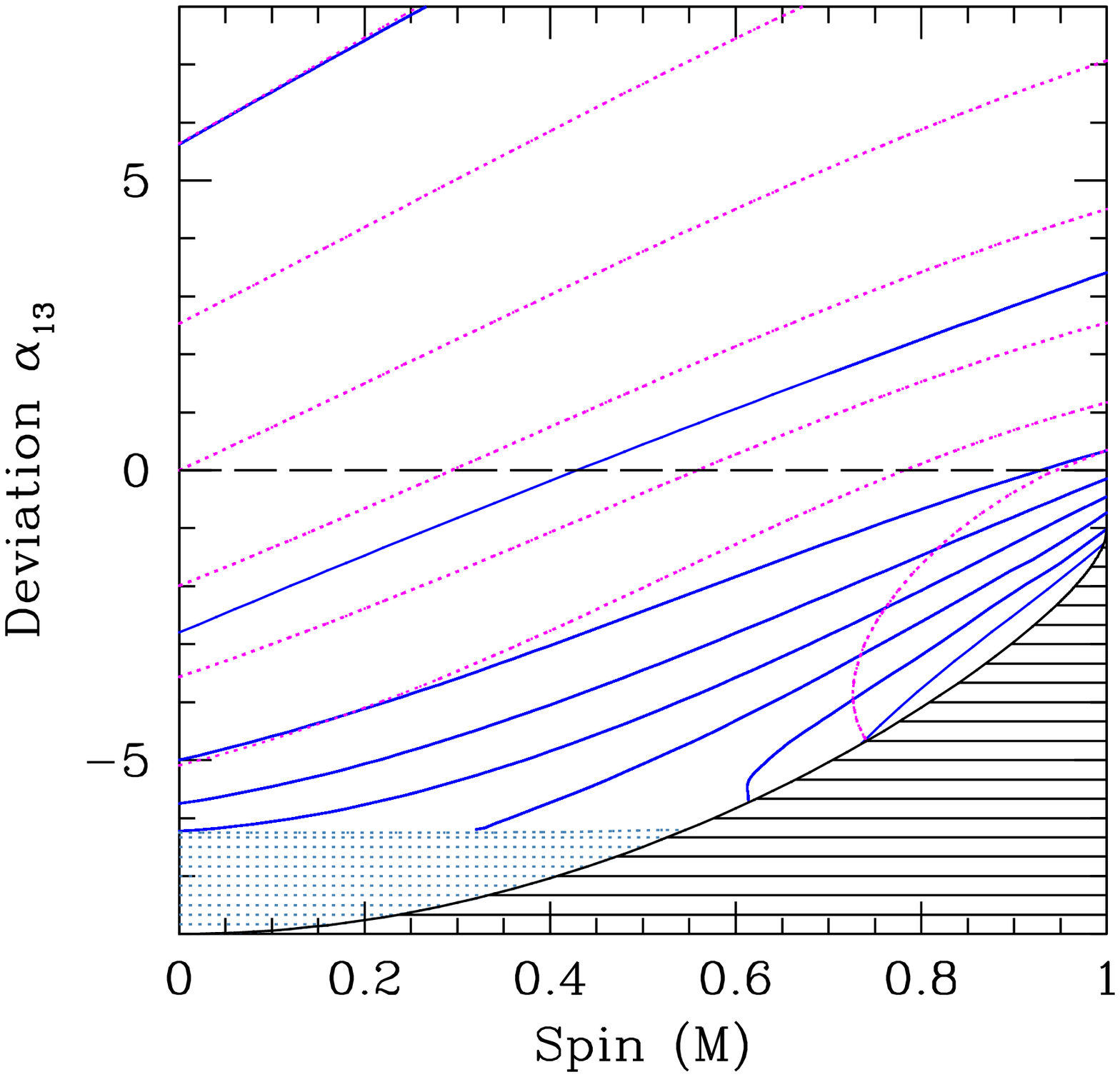,width=0.4\textwidth}
\psfig{figure=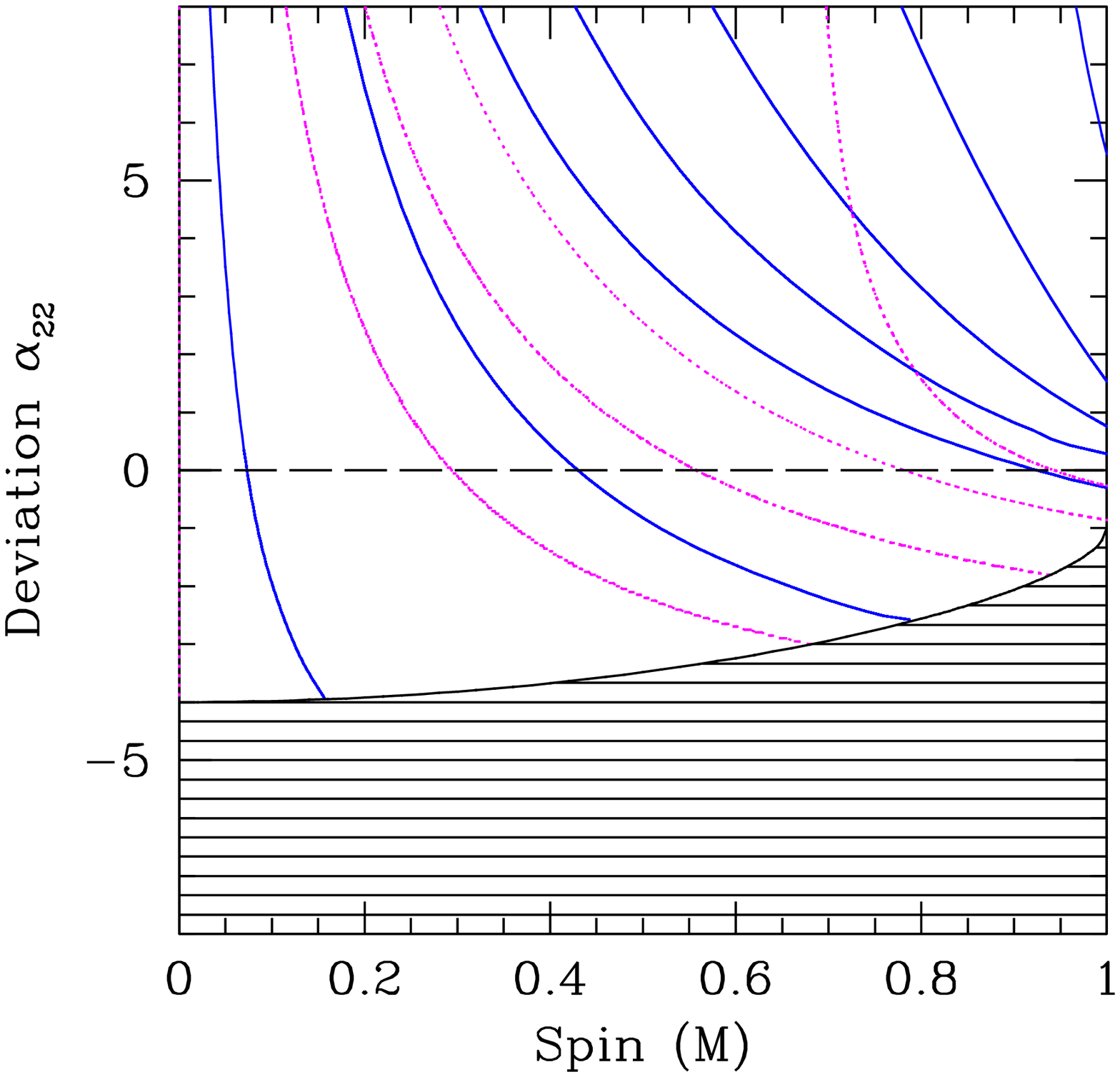,width=0.4\textwidth}
\end{center}
\caption{Contours of constant $g$-mode frequency for a $10M_\odot$ black hole as a function of the spin and the deviation parameters $\alpha_{13}$ (left) and $\alpha_{22}$ (right). In the top panel, frequency contours are shown as solid blue curves with frequencies (top to bottom) $50~{\rm Hz}$, $100~{\rm Hz}$, $200~{\rm Hz}$, $300~{\rm Hz}$, $400~{\rm Hz}$, $500~{\rm Hz}$, $600~{\rm Hz}$, and $700~{\rm Hz}$. For reference, contours of constant ISCO radius are shown as dotted magenta lines with radii (top to bottom) $8M$, $7M$, $\ldots$, $2M$. In the gray dotted region, the radial epicyclic frequency does not vanish at the ISCO and $g$-modes may not exist. In the bottom panel, frequency contours are shown as solid blue curves with frequencies (left to right) $75~{\rm Hz}$, $100~{\rm Hz}$, $200~{\rm Hz}$, $300~{\rm Hz}$, $400~{\rm Hz}$, $500~{\rm Hz}$, and $600~{\rm Hz}$. Contours of constant ISCO radius are shown as dotted magenta lines with radii (left to right) $6M$, $5M$, $\ldots$, $2M$. In both cases, the contours of constant $g$-mode frequency are mostly aligned with the contours of constant ISCO radius except for larger deviations from the Kerr metric if the ISCO lies at a radius smaller than roughly $3M$. The black shaded region marks the excluded part of the parameter space.}
\label{fig:gmode}
\end{figure*}

\begin{figure*}[ht]
\begin{center}
\psfig{figure=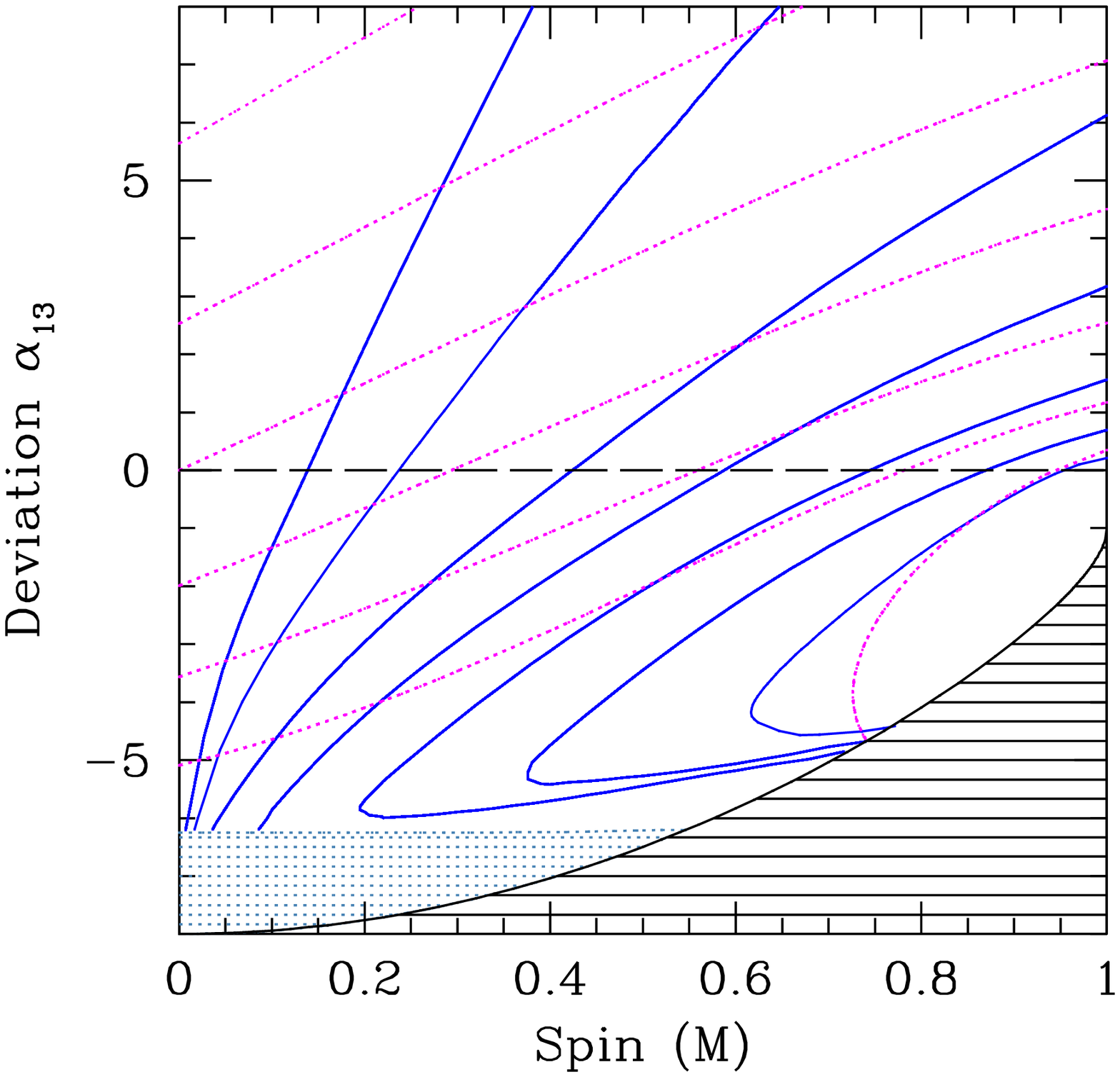,width=0.4\textwidth}
\psfig{figure=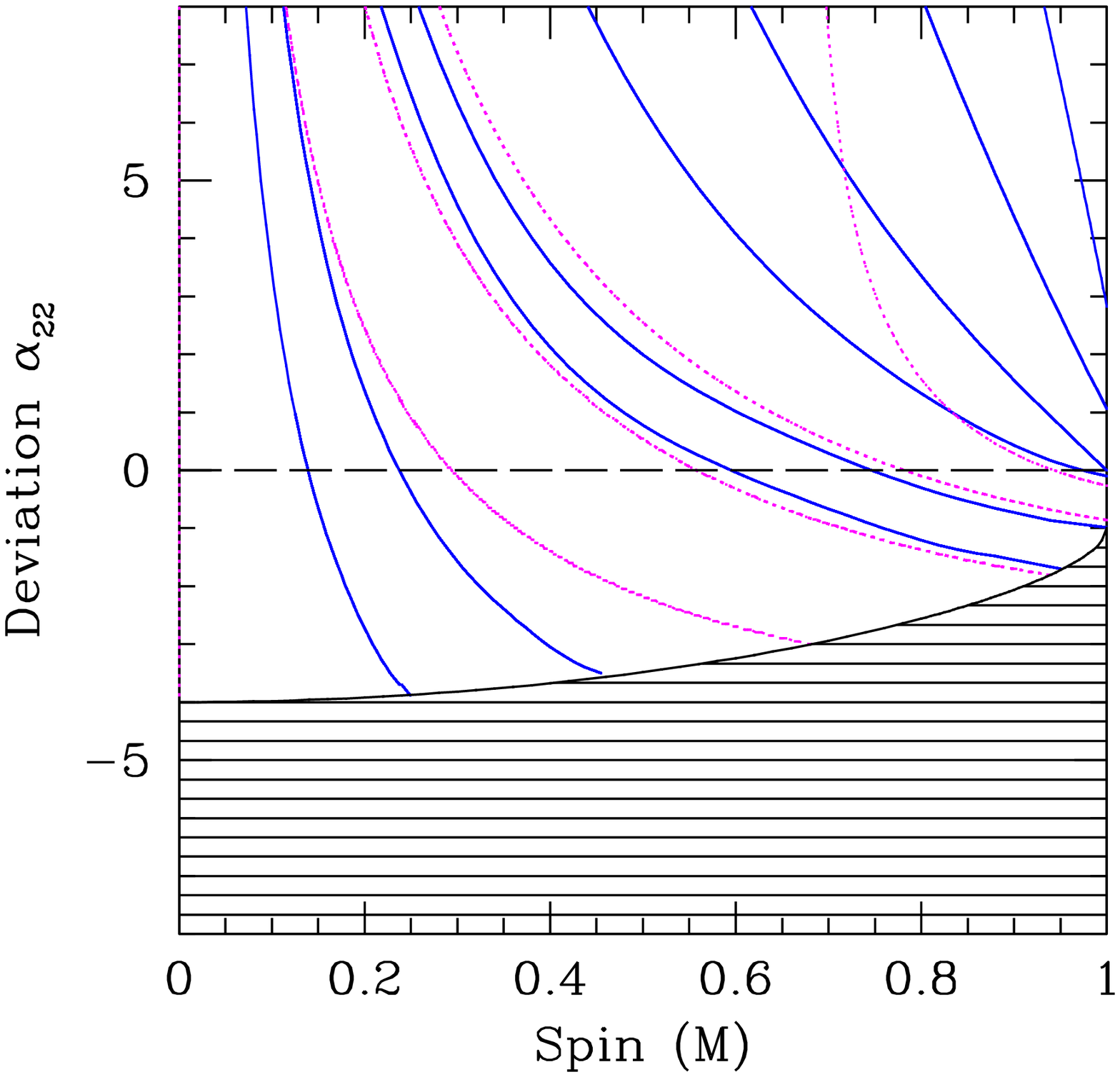,width=0.4\textwidth}
\end{center}
\caption{Contours of constant $c$-mode frequency for a $10M_\odot$ black hole as a function of the spin and the deviation parameters $\alpha_{13}$ (left) and $\alpha_{22}$ (right). In the top panel, frequency contours are shown as solid blue curves with frequencies (left to right) $5~{\rm Hz}$, $10~{\rm Hz}$, $25~{\rm Hz}$, $50~{\rm Hz}$, $100~{\rm Hz}$, $200~{\rm Hz}$, and $400~{\rm Hz}$. For reference, contours of constant ISCO radius are shown as dotted magenta lines with radii (top to bottom) $8M$, $7M$, $\ldots$, $2M$. The $c$-mode contours are only roughly aligned with the ISCO contours for large values of the spin and for deviations $|\alpha_{13}|\lesssim3$. In the gray dotted region, the radial epicyclic frequency does not vanish at the ISCO and $c$-modes may not exist. In the bottom panel, frequency contours are shown as solid blue curves with frequencies (left to right) $5~{\rm Hz}$, $10~{\rm Hz}$, $50~{\rm Hz}$, $100~{\rm Hz}$, $500~{\rm Hz}$, $1000~{\rm Hz}$, $2000~{\rm Hz}$, and $4000~{\rm Hz}$. Contours of constant ISCO radius are shown as dotted magenta lines with radii (left to right) $6M$, $5M$, $\ldots$, $2M$. The contours of constant $c$-mode frequency are mostly aligned with the contours of constant ISCO radius except for larger deviations from the Kerr metric if the ISCO lies at a radius smaller than roughly $3M$. The black shaded region marks the excluded part of the parameter space.}
\label{fig:cmode}
\end{figure*}

\begin{figure*}[ht]
\begin{center}
\psfig{figure=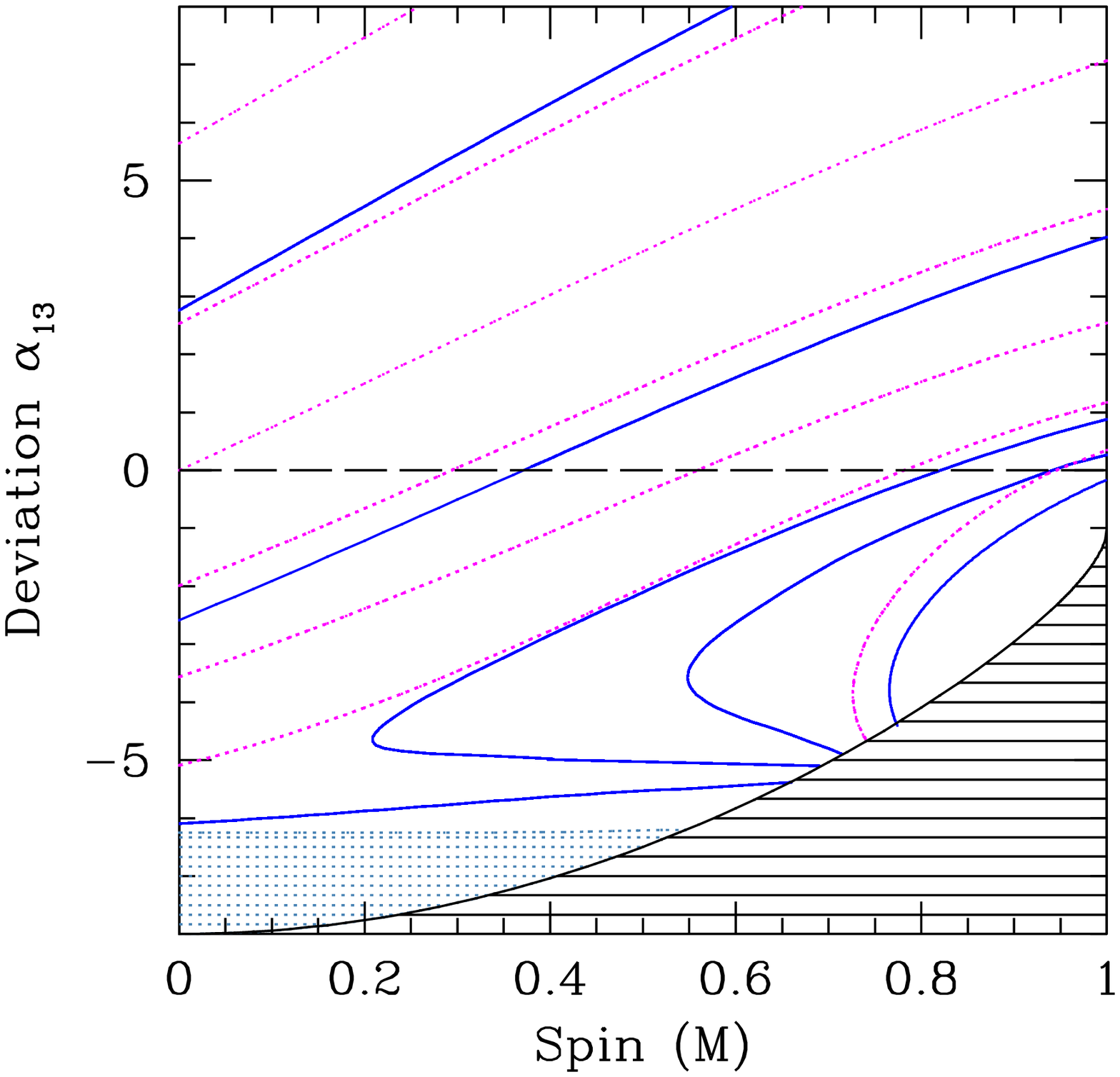,width=0.4\textwidth}
\psfig{figure=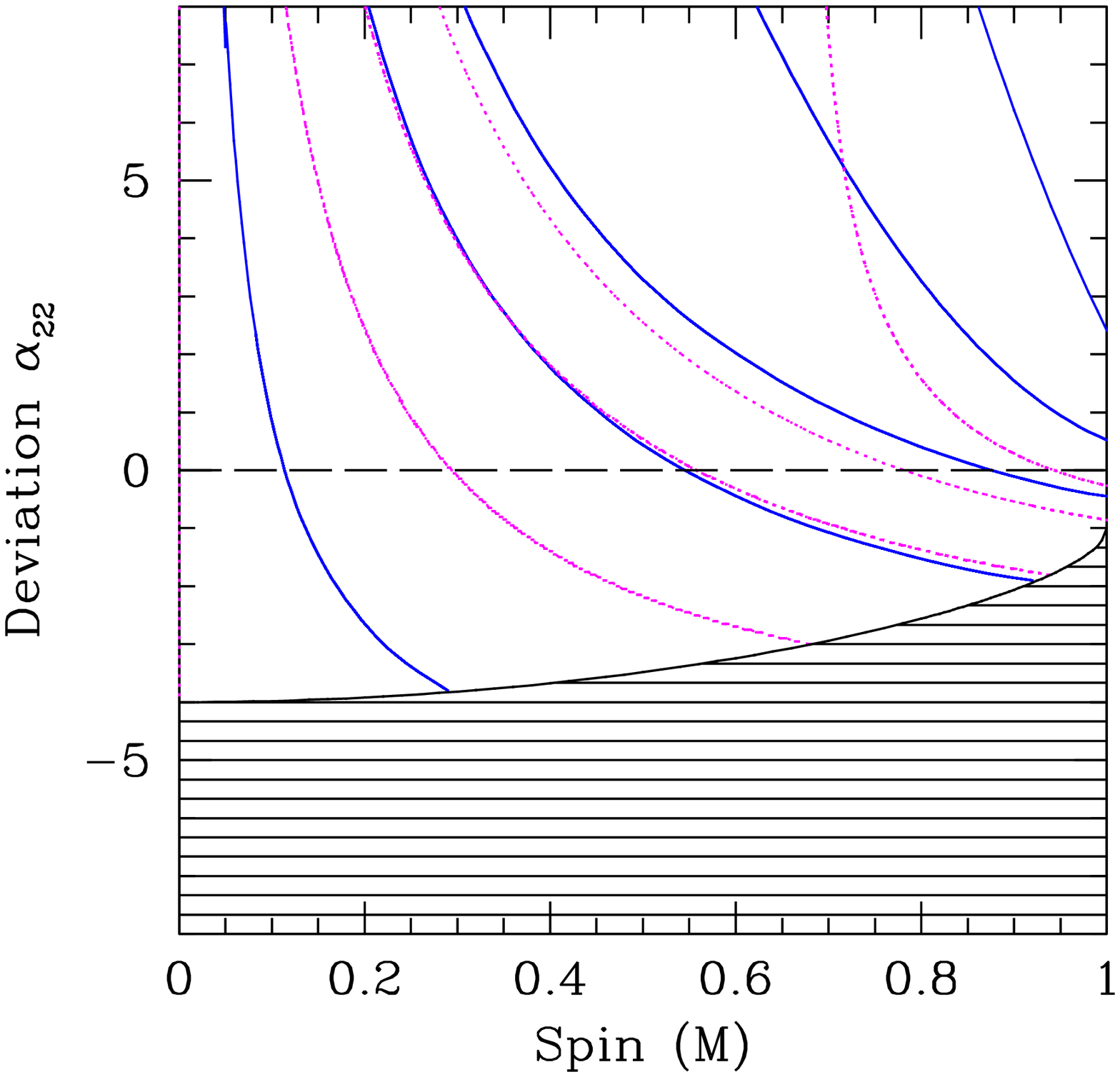,width=0.4\textwidth}
\end{center}
\caption{Contours of constant Keplerian frequency for a $3:1$ resonance between the Keplerian and radial epicyclic frequencies for a $10M_\odot$ black hole as a function of the spin and the deviation parameters $\alpha_{13}$ (left) and $\alpha_{22}$ (right). In the top panel, frequency contours are shown as solid blue curves with frequencies (top to bottom) $100~{\rm Hz}$, $150~{\rm Hz}$, $250~{\rm Hz}$, $450~{\rm Hz}$, $600~{\rm Hz}$, and $900~{\rm Hz}$. In the gray dotted region, the radial epicyclic frequency does not vanish at the ISCO and resonance modes may not exist. The additional blue line at the bottom forms the disjoint second branch of the $250~{\rm Hz}$ contour which emerges at the boundary of the gray dotted region at a value of the spin $a\approx-0.2M$ (not shown). For reference, contours of constant ISCO radius are shown as dotted magenta lines with radii (top to bottom) $8M$, $7M$, $\ldots$, $2M$. In the bottom panel, frequency contours are shown as solid blue curves with frequencies (left to right) $200~{\rm Hz}$, $300~{\rm Hz}$, $500~{\rm Hz}$, $1000~{\rm Hz}$, and $1500~{\rm Hz}$. Contours of constant ISCO radius are shown as dotted magenta lines with radii (left to right) $6M$, $5M$, $\ldots$, $2M$. In both cases, the contours of constant $g$-mode frequency are mostly aligned with the contours of constant ISCO radius except for large values of the parameter $\alpha_{22}$. The black shaded region marks the excluded part of the parameter space.}
\label{fig:res}
\end{figure*}

In addition to the thermal accretion disk spectra and relativistically broadened iron lines, quasiperiodic variability constitutes a potential third approach to measure the spins of black holes and even of potential deviations from the Kerr metric. In the following, I investigate the dependence of QPO frequencies on the spin and the deviation parameters in two particular QPO models, the diskoseismology model (see Ref.~\cite{wag08} for a review) and the resonance model \cite{kluz01,abra03}. Further viable QPO models exist (e.g., \cite{otherQPOmodels}), but I do not consider them here, because these do not predict fixed values of the spin (or the deviation parameters) at a given QPO frequency.

In the diskoseismology model, QPOs can arise as so-called gravity modes ($g$-modes \cite{per97}) and corrugation modes ($c$-modes \cite{sil01}). Here I focus on the fundamental $g$- and $c$-modes. In the resonance model, QPOs can be identified as resonances between the dynamical frequencies which preferentially occur at small-integer frequency ratios. As an example, I focus on a $3:1$ resonance between the Keplerian and radial epicyclic frequencies; other resonances should have a very similar dependence of the deviation parameters. 

Expressions for the Keplerian as well as the radial and vertical epicyclic frequencies for particles on circular equatorial orbits in the metric given by Eq.~(\ref{eq:metric}) can be found in Ref.~\cite{joh13metric}. The fundamental $g$-mode occurs at the radius where the radial epicyclic frequency reaches its maximum and the fundamental $c$-mode corresponds to the Lense-Thirring frequency (i.e., the difference between the Keplerian and vertical epicyclic frequencies) at the ISCO \cite{per97,sil01}. A parametric resonance between the dynamical frequencies usually occurs at a different radius, which is determined directly by the frequency ratio of the resonance \cite{kluz01,abra03}.

In Figs.~\ref{fig:gmode}--\ref{fig:res}, I plot contours of the fundamental $g$- and $c$-modes and of the Keplerian frequency in the $3:1$ resonance, respectively. For reference, I likewise plot contours of constant ISCO radius. The contours of the $g$-mode and resonance frequencies are mostly aligned with the ISCO contours except for very high frequencies. The contours of the $c$-modes are less aligned with the ISCO contours especially for very low frequencies in the case of deviations described by the parameter $\alpha_{13}$.

The dependence of these QPO modes on the spin and the deviation parameters is similar to the one in the quasi-Kerr metric and the metric of Ref.~\cite{JPmetric} as discussed in Refs.~\cite{joh11a,jp13,bambiqpo}. However, since at the lowest order in the deviation functions the metric in Eq.~(\ref{eq:metric}) contains four deviation parameters instead of only one as in the other two metrics, the departure of the frequency contours from the contours of constant ISCO radius are generally larger.

Since the dynamical frequencies and the location of the ISCO depend only marginally on the deviation parameter $\epsilon_3$ and not at all on the deviation parameter $\alpha_{52}$ (apart from the radial epicyclic frequency, which is affected only slightly), their effect on the $g$-, $c$-, and resonance modes are small.

%%%%%%%%%%%%%%%%%%%%%%%%%%%%%%%%%%%%%%%
\section{Conclusions}
\label{sec:conclusions}

In this paper, I analyzed three potential high-energy probes of deviations from the Kerr metric parametrized by a new Kerr-like metric \cite{joh13metric}: thermal continuum spectra, relativistically broadened iron lines, and quasiperiodic variability. The Kerr-like metric describes a black hole and depends on four deviation parameters which affect different components of the metric. This work extends previous analyses that focused on other Kerr-like metrics \cite{GB06,JPmetric} which depend on only one deviation parameter and generally harbor naked singularities (see Ref.~\cite{joh13pathols}).

I showed that all three observables depend significantly on two deviation parameters ($\alpha_{13}$ and $\alpha_{22}$) which modify the $(t,t)$, $(t,\phi)$, and $(\phi,\phi)$ elements. I also showed that in the case of limb darkenend emission a third deviation parameter, $\epsilon_3$, which alters all nonvanishing metric elements affects both the thermal spectra and iron line profiles, while a fourth deviation parameter, $\alpha_{52}$, which modifies only the $(r,r)$ component affects only the iron line profiles at higher inclinations. The latter two parameters have a relatively small effect on QPO frequencies and only lead to potentially detectable frequency shifts if their magnitudes are very large.

\begin{table*}[ht]
\begin{center}
\footnotesize
\begin{tabular}{lcccccccc}
\multicolumn{9}{c}{}\\
\hline \hline

Parameter     & ISCO Radius & Thermal Spectrum & Thermal Spectrum & Iron Line   & Iron Line          & $g$-Mode & $c$-Mode & Resonance \\
              &          & (isotropic)      & (limb darkened)  & (isotropic) & (limb darkened)    &          &          &           \\
\hline
$\epsilon_3$  & Moderate & Weak            & Strong           & Weak       & Moderate${\rm ^a}$ & Weak    & Weak    & Weak     \\
$\alpha_{13}$ & Strong   & Strong          & Strong           & Strong      & Strong             & Strong   & Strong   & Strong    \\
$\alpha_{22}$ & Strong   & Strong          & Strong           & Strong      & Strong             & Strong   & Strong   & Strong    \\
$\alpha_{52}$ & None     & Weak            & Weak            & Weak       & Moderate${\rm ^a}$ & Weak    & None     & Weak     \\
\hline
\end{tabular}
\end{center}
\footnotetext{High disk inclinations only.}
\caption{Effects of the deviation parameters $\epsilon_3$, $\alpha_{13}$, $\alpha_{22}$, and $\alpha_{52}$ on the location of the ISCO as well as on thermal accretion disk spectra, relativistically broadened iron lines, and QPOs. The parameters $\alpha_{13}$ and $\alpha_{22}$ have a strong effect on all quantities. In the case of thermal disk spectra and iron lines, the effect of the parameters $\epsilon_3$ and $\alpha_{52}$ depends on the emission type of the accretion disk (isotropic or limb darkened). The location of the ISCO radius is affected only slightly by the parameter $\epsilon_3$ and is independent of the parameter $\alpha_{52}$. These two parameters likewise have a marginal effect on the QPOs.
}
\end{table*}

Since the ISCO of the spacetime depends on the spin and the parameters $\alpha_{13}$, $\alpha_{22}$, and $\epsilon_3$, it is important to characterize the non-Kerr effects on the three observables for those parameter combinations that belong to the same ISCO radius. I showed that thermal spectra and iron line profiles are practically indistinguishable if the ISCO lies at a sufficiently large radius corresponding to small to intermediate spin values of Kerr black holes. For Kerr black holes with high spins, however, both the thermal spectra and iron line profiles differ for sets of parameters with the same ISCO radius. This difference manifests already for small values of these deviation parameters in the case of the thermal spectra and for larger values of these parameters in the case of the iron lines unless the emissivity index is very high. Such a difference is qualitatively new and has not been seen in other Kerr-like metrics (cf. e.g., \cite{jp13}).

I also showed that in the diskoseismology and resonance models, QPO frequencies likewise depend strongly on the deviation parameters $\alpha_{13}$ and $\alpha_{22}$ and only marginally on the other two. In both models, contours of constant frequencies are mostly aligned with the contours of constant ISCO radius with exceptions at high spins and for low-frequency $c$-modes. In Table~I, I summarize the effects of the deviation parameters on the various observables.

\begin{figure*}[ht]
\begin{center}
\psfig{figure=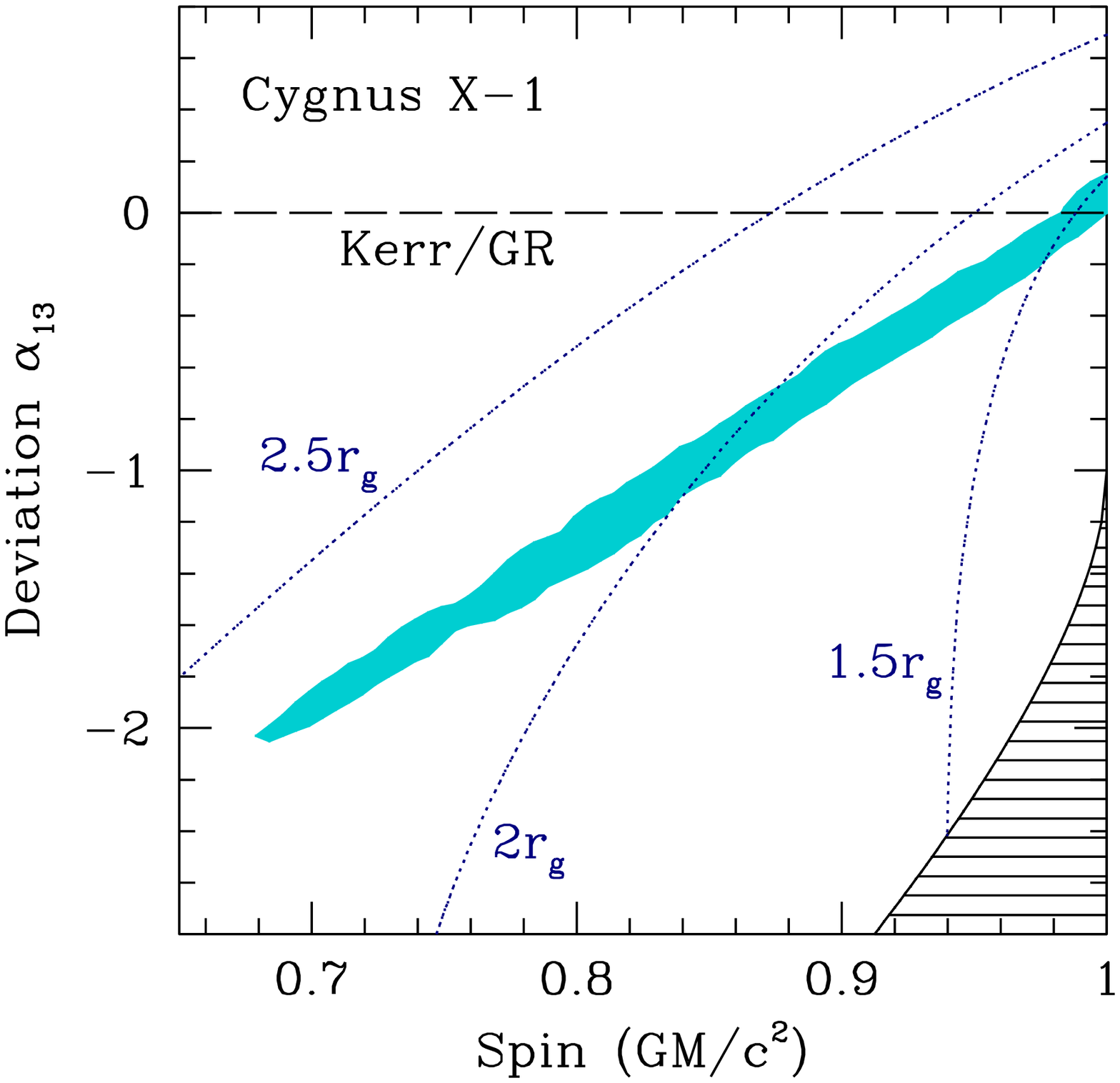,width=0.4\textwidth}
\psfig{figure=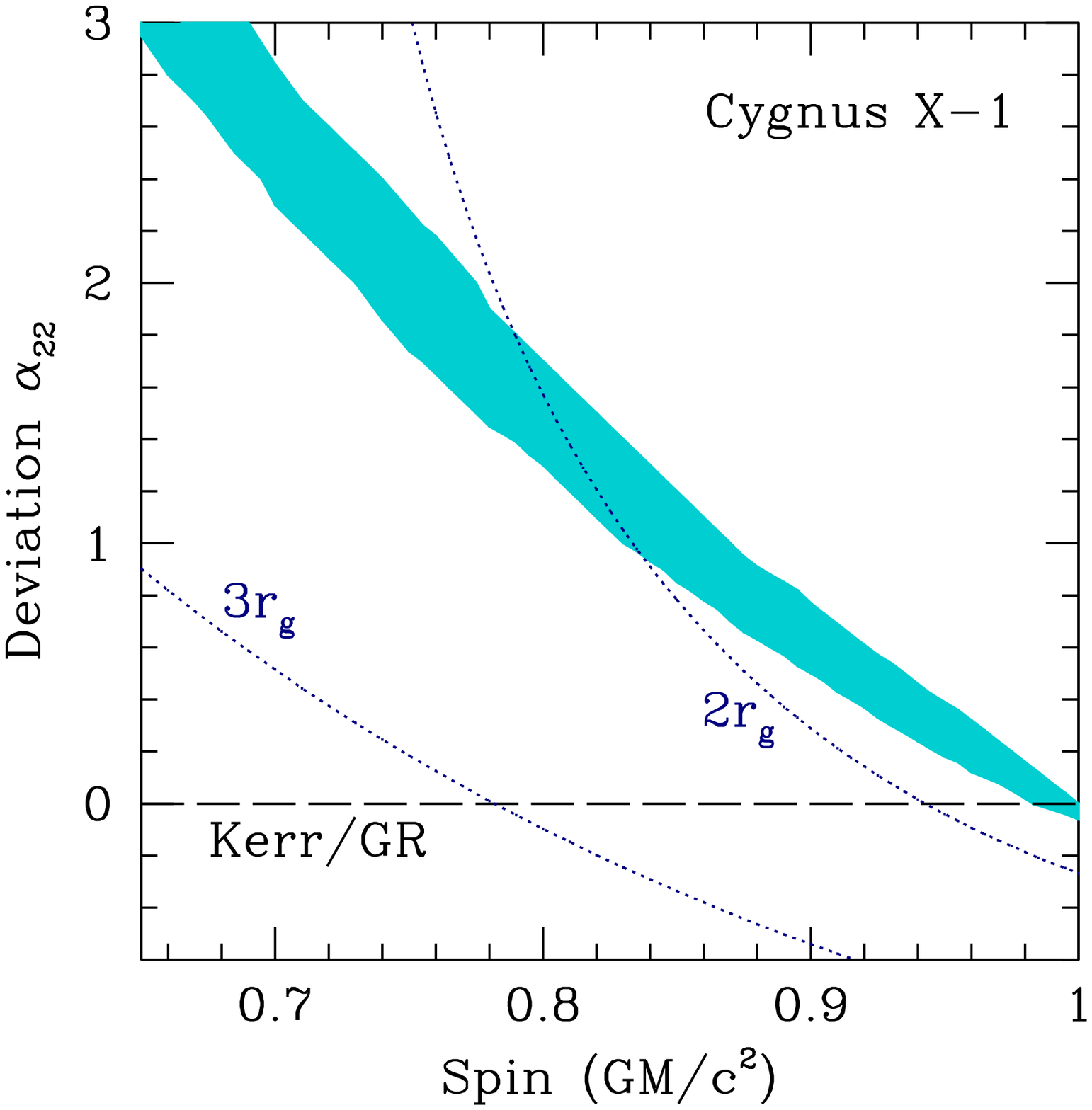,width=0.4\textwidth}
\end{center}
\caption{Rough constraints (light blue regions) on the spin $a$ and deviation parameters $\alpha_{13}$ (left) and $\alpha_{22}$ (right) from the stellar-mass black hole Cygnus~X--1. The light blue regions show the range of thermal spectra as a function of the spin and the respective deviation parameter for which the flux density at energies $0.7~{\rm keV} \leq E \leq4.0~{\rm keV}$ lies between the flux densities of the spectra of Kerr black holes with spins $a=0.983M$ and $a=M$ corresponding to the $3\sigma$ limits of the recent spin measurement by Gou et al.~\cite{Gou14}. The other system parameters are held fixed at their values given in Ref.~\cite{Gou14}. In both panels, the allowed range of spectra is a narrow region that does not lie along lines of constant ISCO radius (blue dotted lines), because at these high spin values the relativistic effects in the innermost region of the accretion disk affect the spectra significantly. In the left panel, the spin and the parameter $\alpha_{13}$ are constrained to the approximate intervals $a>0.68M$ and $-2.1<\alpha_{13}<0.2$, while, in the right panel, the constrained region is bound only by the lower limit $\alpha_{22}>-0.1$ and extends to small values of the spin and very large values of the deviation parameter $\alpha_{22}$ (shown up to a value $\alpha_{22}=3$). The black long-dashed line labeled ``Kerr/GR'' corresponds to a Kerr black hole. The black shaded region in the left panel marks the excluded part of the parameter space.}
\label{fig:cygx1}
\end{figure*}

Already, a wealth of x-ray data has been collected which led to a number of spin measurements with the continuum fitting and iron line techniques (see, e.g., Refs.~\cite{rey13,mccl13}). In order to constrain potential deviations from the Kerr metric, the ability to unambiguously identify such deviations is essential. Unfortunately, all observables discussed in this paper strongly correlate with the location of the ISCO at least in the case of black holes with small to intermediate spins. Therefore, these techniques will likely only be able to constrain combinations of values of the spin and the deviation parameters which roughly correspond to the same ISCO radius unless the black hole has a high spin. The consideration of additional deviation parameters, in the form of either higher order terms in the deviation functions given by Eqs.~(\ref{eq:A1})--(\ref{eq:f}) or generalized metrics which contain more than four such functions, would most likely only increase the level of degeneracy between spacetimes with similar ISCO radii but with different values of the spin and the deviation parameters. Nonetheless, an analysis of such observations in the Kerr-like background should be able to place first constraints on deviations from the Kerr metric. For some sources, two or even three of the above observables have been detected and a combination of these measurements may further improve the constraints.

Reference~\cite{jp13} estimated the precision required of future x-ray missions such as Astro-H and ATHENA+ to measure deviations from the Kerr metric parametrized by the metric of Ref.~\cite{JPmetric} with iron lines. They found that black holes with higher spin values allow for tighter constraints on the deviation parameter and that a precision of $\sim5\%$ leads to constraints on the deviation parameter to less than order unity for spin values $|a|\gtrsim0.5M$. Since the parameters $\alpha_{13}$ and $\alpha_{22}$ in the metric of Ref.~\cite{joh13metric} affect the line profiles in roughly the same manner (by magnitude) as the deviation parameter in the metric of Ref.~\cite{JPmetric}, these requirements should be similar if the former metric is used as the underlying spacetime.

In the case of thermal disk spectra, the Galactic black hole Cygnus X--1 is a prime candidate for a constraint on potential deviations from the Kerr metric. Assuming that this object is a Kerr black hole, Gou et al.~\cite{Gou14} recently measured a spin value $a>0.983M$ at $3\sigma$ confidence. In order to find a very rough estimate of such a constraint, I determined graphically the range of spectra as a function of the spin and the parameters $\alpha_{13}$ and $\alpha_{22}$ for which the flux density lies between the flux densities of the spectra of Kerr black holes with spins $a=0.983M$ and $a=M$ at energies $0.7~{\rm keV} \leq E \leq4.0~{\rm keV}$. In all cases, I considered only one deviation parameter and set the other deviation parameters to zero. The remaining system parameters were held fixed at their values given in Ref.~\cite{Gou14} (inclination $i=27.1^{\circ}$, mass $M=14.8~{\rm M_\odot}$, distance $D=1.86~{\rm kpc}$, spectral hardening factor $f_{\rm col}=1.6$). Since the normalization of the spectra is arbitrary in this analysis, I simply set the mass accretion rate to a constant value $\dot{M}=10^{19}~{\rm g~s^{-1}}$. All spectra assume limb darkened disk emission.

This procedure results in two fairly narrow ranges of allowable values in the $(a,\alpha_{13})$ and $(a,\alpha_{22})$ parameter spaces, respectively. I plot these regions in Fig.~\ref{fig:cygx1}. Note that these regions do not lie along lines of constant ISCO radius, because the relativistic effects in the innermost part of the accretion disk are important at such high spin values. I obtain the approximate constraints $a>0.68M$ and $-2.1<\alpha_{13}<0.2$ in the case of deviations described by the parameter $\alpha_{13}$ as well as $\alpha_{22}>-0.1$ in the case of deviations described by the parameter $\alpha_{22}$. In the latter case, the allowable range of spectra extends to low values of the spin and very large values of the deviation parameter $\alpha_{22}$ (not shown).

A measurement of the parameters $\epsilon_3$ or $\alpha_{52}$ with either thermal spectra or iron line profiles will be much more complicated, because their overall effect on the spectra is small. Other observables such as x-ray polarization \cite{schnitt,krawc12} as well as time-resolved iron line profiles and time lags \cite{lags} may constitute additional probes of deviations from the Kerr metric.

A different approach is the imaging of the shadows of the Galatic center black hole Sgr~A* and of the supermassive black hole in M87 with the Event Horizon Telescope \cite{EHT}. Reference~\cite{joh13rings} showed that the shapes of the shadows of supermassive black holes described by the metric in Eq.~(\ref{eq:metric}) depend only on the spin and the deviation parameters $\alpha_{13}$ and $\alpha_{22}$. Consequently, using either x-ray or (sub)mm observations, deviations described by the parameters $\alpha_{13}$ and $\alpha_{22}$ will probably be easier to detect in practice, at least in the near future.

\acknowledgments

This work was supported by a CITA National Fellowship at the University of Waterloo and in part by Perimeter Institute for Theoretical Physics. Research at Perimeter Institute is supported by the Government of Canada through Industry Canada and by the Province of Ontario through the Ministry of Research and Innovation.

%%%%%%%%%%%%%%%%%%%%%%%%%%%%%%%%%
%\appendix
\appendix
\section{PHOTON EMISSION ANGLE}

Here, I derive an expression for the emission angle $\zeta$ of a given photon momentum vector relative to the disk normal. My derivation follows closely Refs.~\cite{Speith93,Speith95}.

The emission angle of the photon momentum vector is measured in a local inertial frame with a basis $\{e_{\hat t},e_{\hat r},e_{\hat \theta},e_{\hat \phi}\}$. In the equatorial plane, where the accretion disk is located, the disk normal $\hat{n}$ is simply $e_{\hat \theta}$, which is related to the coordinate basis $\{e_t,e_r,e_\theta,e_\phi\}$ of the metric by the equation \cite{joh13rings}
\be
{\hat n}=e_{\hat \theta}=\left.\frac{1}{\sqrt{g_{\theta\theta}}} e_\theta \right|_{\theta=\frac{\pi}{2}}.
\ee
The emission angle is, then, given by the expression
\be
\cos \zeta = \frac{\left|\bar{p}_\perp^{\rm e}\right|}{\left| \bar{p}^{\rm e} \right|} = - \frac{\hat{n}^\alpha p^{\rm e}_\alpha}{u^\alpha p^{\rm e}_\alpha},
\ee
where $p^{\rm e}_\alpha$ is the (contravariant) photon 4-momentum, $\bar{p}^{\rm e}$ the photon 3-momentum, $\bar{p}_\perp^{\rm e}$ its component normal to the disk, and $u^\alpha$ is the 4-velocity of the disk plasma. Since from Eq.~(\ref{eq:thetaEOM})
\be
p^{\rm e}_\theta = \left. \frac{\sqrt{\Theta(\theta)}}{\tilde{\Sigma}} \right|_{\theta=\frac{\pi}{2}} = \frac{\sqrt{\eta}}{\tilde{\Sigma}},
\ee
\be
\hat{n}^\alpha p^{\rm e}_\alpha = \frac{\sqrt{\eta}}{\sqrt{r_{\rm e}^2 + \epsilon_3 \frac{M^3}{r_{\rm e}}}}.
\ee
From Eq.~(\ref{eq:redshift}), the redshift is given by the expression
\be
g = \frac{p_t^{\rm obs}}{u^\alpha p^{\rm e}_\alpha} = -\frac{1}{u^\alpha p^{\rm e}_\alpha}.
\ee
Therefore, the emission angle can be expressed as in Eq. (\ref{eq:coszeta}). In the case of a Kerr black hole, this expression reduces to the equation
\be
\cos\zeta = \frac{g\sqrt{\eta}}{r_{\rm e}}
\ee
as in Refs.~\cite{Speith93,Speith95}.

\end{document}